\newcommand{\Teff}{\mbox{$T_{\rm eff}$}~}
\newcommand{\teff}{\mbox{$T_{\rm eff}$}}
\newcommand{\Lbol}{\mbox{$L_{\rm bol}$}}
\newcommand{\Lsun}{\mbox{$L_{\odot}$}}

\newcommand{\rrg}{\mbox{$R_{\rm RG}$}}

\newcommand{\rcomp}{\mbox{$R_{\rm comp}$}}
\newcommand{\Rsun}{\mbox{$R_{\odot}$~}}
\newcommand{\rsun}{\mbox{$R_{\odot}$}}
\newcommand{\Msun}{\mbox{$M_{\odot}$~}}
\newcommand{\msun}{\mbox{$M_{\odot}$}}

\newcommand{\logg}{log~{\it g}}
\newcommand{\kms}{km s$^{-1}$}

\newcommand{\as}{\mbox{$^{\prime\prime}$~}}

\newcommand{\rv}{$R_V$~}

\newcommand{\beqa}{\begin{eqnarray}} 
\newcommand{\eeqa}{\end{eqnarray}}

\newcommand{\Ebv}{\mbox{$E_{B\!-\!V}$}~}
\newcommand{\ebv}{\mbox{$E_{B\!-\!V}$}}

\newcommand{\wleff}{\mbox{$\lambda_{\rm eff}$}}
\newcommand{\ecmsang}{\mbox{ergs~cm$^{-2}$~s$^{-1}$~\AA$^{-1}$}}

\documentclass[12pt,preprint]{aastex} 
\usepackage{graphicx,pdflscape}
\pdfoutput=1

\slugcomment{ApJ, in press; DOI: 10.3847/1538-4357/ad712f }

\shorttitle{A red giant$+$subgiant binary  }
\shortauthors{Bianchi et al.}

\begin{document}

\title{Revealing the elusive companion of the red-giant binary 2MASSJ05215658+4359220 from UV HST and Astrosat/UVIT data.} 

\author{Luciana Bianchi\altaffilmark{1},  John  Hutchings\altaffilmark{3}, Ralph Bohlin\altaffilmark{2},
 David Thilker\altaffilmark{1}, Emanuele Berti\altaffilmark{1} } 
\altaffiltext{1}{Dept. of Physics \& Astronomy, The Johns Hopkins University, 3400 N. Charles St.,  Baltimore, MD 21218, USA; 
http://dolomiti.pha.jhu.edu}
\altaffiltext{2}{Space Telescope Science Institute, 3400 San Martin Dr., Baltimore, MD 21218, USA} 
\altaffiltext{3}{Herzberg Institute for Astrophysics, Victoria, CA}
\email{bianchi@jhu.edu}

\begin{abstract}
  Black hole demographics in different environments is critical in view of recent results on massive-stars binarity, and of the multi-messenger detectability of compact objects mergers. But the identification and characterization of non-interacting black holes is elusive, especially in the sparse field stellar population.  A candidate  non-interactive black-hole~(BH)~+~red-giant~(RG) binary system,  2MASSJ05215658+4359220, was identified by \citet{Thompsonetal2019}.
We obtained Astrosat/UVIT Far-Ultraviolet (FUV) imaging and Hubble Space Telescope (HST)  UV$-$optical imaging and spectroscopy of the source,  to test possible scenarios for the optically-elusive companion. 
HST/STIS spectra from $\approx$1,600 to 10,230\AA~  are best fit 
 by the combination of  two stellar sources, a red giant with \Teff=4250$\pm$150~K, \logg=2.0, \rrg$\sim$27.8\rsun (assuming a single-temperature atmosphere), and a subgiant companion with \Teff=6,000K, \rcomp=2.7\rsun,  or \teff=5,270K, \rcomp=4.2\Rsun using models with one-tenth or one-third solar metallicity respectively, \logg=3.0, 
 extinction \ebv=0.50$\pm$0.2,  adopting the DR3 Gaia distance D=2463$\pm$120~pc. 
 No FUV data existed prior to our programs. STIS spectra give an upper limit of 10$^{-17}$\ecmsang ~shortwards of 2300\AA; an upper limit of 
 $\gtrsim$25.7~ABmag was obtained in two UVIT FUV broad-bands.  The non-detection of FUV flux rules out 
 a compact companion such as a hot WD. 
The STIS spectrum shows strong MgII~$\lambda$2800\AA~  emission, typical of chromospherically active red giants. The masses inferred by comparison with evolutionary tracks, $\sim$1\Msun for the red giant and between 1.1-1.6\Msun for the  subgiant companion, suggest past mass transfer, although the red giant currently does not fill its Roche lobe.    WFC3 imaging in F218W, F275W, F336W, F475W, and F606W  shows an unresolved source in all filters.
\end{abstract}


\section{Introduction. Non-interacting BHs are theoretically predicted, but elusive.}
\label{s_intro}
Finding and characterizing non-interacting stellar-mass black holes (BHs) is critical  in view  of the 
 capability to detect multi-messenger signals from compact object mergers (e.g., \citet{abbott2017a,abbott2017b,abbott2017c,PostnovYungelson2006}),
and, on the other hand, of the growing evidence  that the majority of massive stars are formed in binaries and that $\sim$70\% of these binaries interact, with 20-30\% of them  possibly merging 
(e.g., \citet{Patrick2019,Sanaetal2014,Sanaetal2017,MoediStefano2017,abadie2010}). These findings challenge our understanding of stellar evolution and stellar populations (e.g., \citet{abbott2016}). 
A missing critical ingredient in this scenario
 is an observational characterization of BH demographics: interacting and merging systems are a biased subset of the whole population, and mass measurements are currently obtained almost exclusively for pulsars and accreting binary systems selected from radio, X-ray, and gamma-ray surveys (e.g., \citet{Ozel2010,Ozel2012,Antoniadis2016}). 
Numerous studies focus on globular clusters (e.g., \citet{banerjee18,dorazio18}),  a special environment where dynamical interactions can alter  native statistics,  that is not representative of the stellar ``field'' and the young populations. For example, \citet{Langer2020} binary evolution models 
predict over 100 OBstar$-$BH systems in the LMC, whereas only one (interacting, LMC X-1) is known.

\section{A wide binary with an optically-elusive companion.}
\label{s_bhrg}

A strong candidate binary  comprised of a red giant and an unseen companion with inferred mass in the stellar BH regime was identified by  \citet{Thompsonetal2019} 
(hereafter T2019) from a large ($\sim$200) APOGEE stellar sample with high radial velocities. 
T2019 reported the red giant to vary in radial velocity (amplitude of 44.6~\kms) and optical brightness (V-mag amplitude of $\sim$0.2mag) with a period of $P$=82.2$\pm$2.5~days, and  interpreted  the synchronous light variation as due to spots in a locked rotation-orbital motion. 
From the optical-IR SED, T2019 derived  \Teff=4530$\pm$89K, \ebv=0.445$\pm$0.050, log(\Lbol/\Lsun)=2.52$\pm$0.03, \logg=2.5, M$\sim$3.2$\pm$1\Msun for the red giant,
 using  a distance of $D$=3106pc from Gaia DR2 corrected for binary motion.   
Gaia DR3 considerably revised this source's parallax, from
$p$=0.322$^{+0.075}_{-0.069}$mas in DR2 to 0.399626$\pm$0.015740~mas  in DR3, bringing the distance 
to  $D$=2.46$\pm$0.12~kpc (as corrected by \citet{elbadry24}).   
  
 In T2019's comprehensive SED analysis, that spans a wide wavelength range,  
  one data point remained inconsistent: a detection of the source in near-UV (NUV), measured by both GALEX NUV and Swift UVOT.UVM2.  T2019 showed that the archival NUV emission is  incompatible with the red-giant SED, as expected, but also with an overall fit of the optical-IR SED if a main-sequence stellar companion were added to account for the NUV flux (see their Figure S8).  Such conclusion, and the inferred companion mass, led  T2019 to suggest a BH companion.

The source is included in  two GALEX broad-band NUV images,
with mag$_{NUV}$ =21.49$\pm$0.36 and =21.44$\pm$0.59 in the AB system.  We examined the GALEX data and confirmed the identification; we also dismiss a possible association of the GALEX NUV source with a very faint Gaia DR3 source at $\sim$3\as away as unlikely.
Five measurements with Swift UVOT.UVM2 (NUV range)  were discussed by T2019. 
Archival NUV data are listed in Table \ref{t_swiftgalexdata}.
No far-UV (FUV) observations existed prior to our programs, presented in this work.

\section{Unveiling the Elusive Companion and the source of the NUV flux} 
\label{s_nuv}

Given the relevance of a  strong candidate for a binary system with a {\it non-interacting}  BH companion, we felt it of great interest to test any possible alternative scenario which could explain the existing ample data; were alternatives  ruled out, the BH hypothesis would gain strength. 
To this aim, we focused on the only unexplained measurements in T2019 SED analysis: the NUV flux, detected consistently by two instruments. We considered the possible scenarios below, and obtained new data from far-UV to optical wavelengths critical to conclusively test all of them.

\subsection{Could a hot white dwarf (WD) or WD pair explain the NUV excess?} 
\label{s_wd}

\subsubsection{Model predictions}
\label{s_wdmodels}

\begin{figure}[!ht]
\center{
\includegraphics[scale=0.16,angle=0]{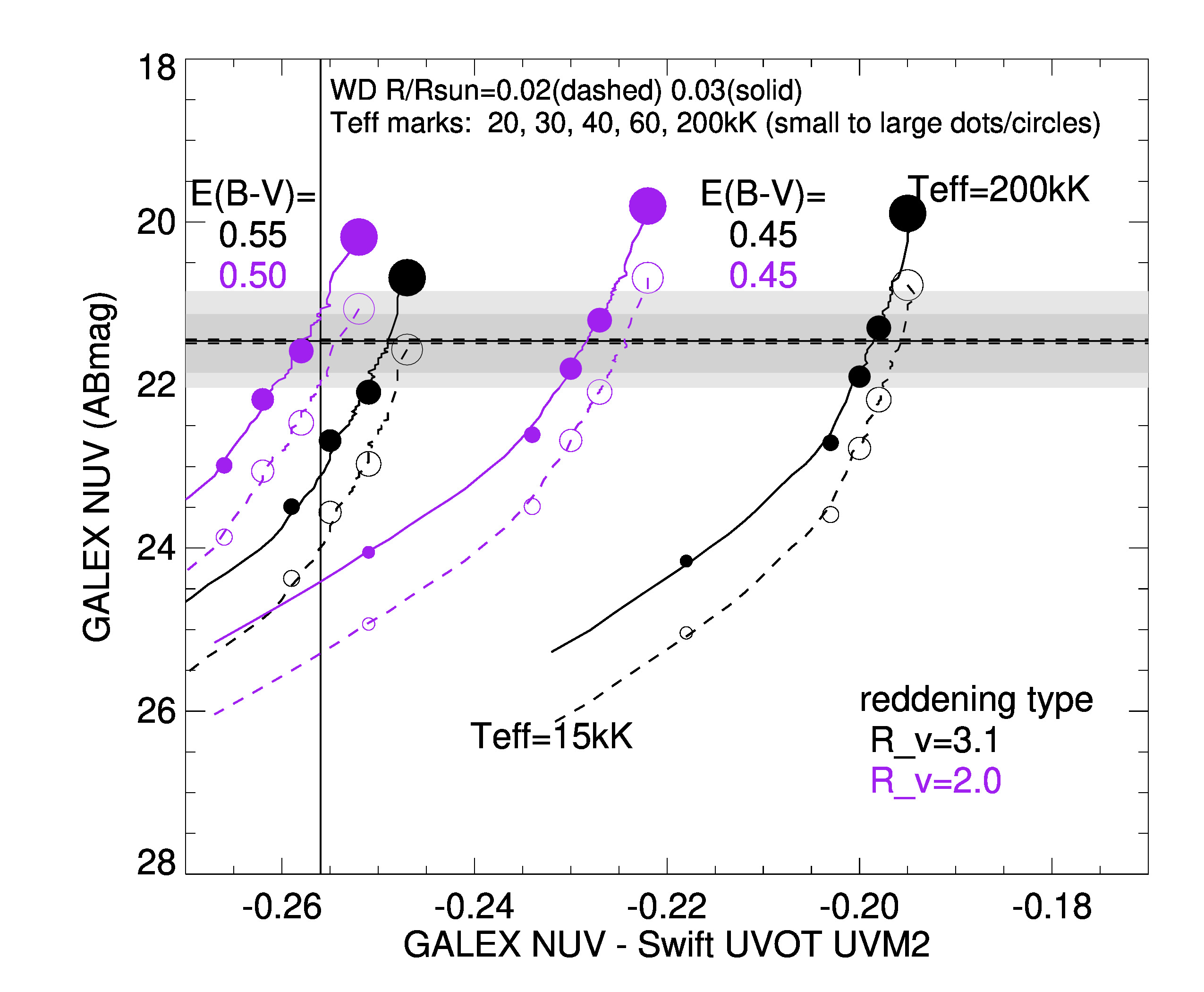}
}
\caption{\footnotesize
{Model magnitude-color sequences for a WD of \Teff=15kK to 200kK 
  with radii of R=0.03 and 0.02\Rsun~ 
  at the distance of the target.   A pair of identical  WDs would be shifted upwards correspondingly to twice the flux, i.e. by 0.75mag. 
  The target's measured GALEX NUV  mags  are shown with horizontal dashed lines (gray shadowed areas show the uncertainties),  
and the NUV$-$UVM2 color with a vertical line.  
Model magnitudes are reddened with two  different extinction types (\rv=3.1, black;  \rv=2, purple  sequences) and different amounts of extinction, for illustration purposes. 
The plot shows that (i) 
  a WD or WD pair (or a cooler main sequence star, not shown) could account for the   GALEX and Swift NUV measurements, unexplained by a red-giant SED (ii)  from NUV-only data,  \Teff  is degenerate with  radius and \ebv;   
 therefore, FUV imaging and a UV$-$optical spectrum were obtained to test the presence of a hot companion and to constrain its parameters if present.} 
\label{f_models}
}
\end{figure}

Among alternative explanations for the detected NUV flux, and possible scenarios for the optically-elusive companion, we had estimated  that both GALEX and Swift NUV fluxes could be compatible with a hot white dwarf (WD), or a WD close pair orbiting the RG at a larger separation  (Figure \ref{f_models}). Their presence would not alter the red-giant optical-IR SED fit by T2019, but rule out the BH companion.  To test such possibility we planned deep observations in FUV with  Astrosat/UVIT \citep{Tandon2020} and the Hubble Space Telescope (HST). 

We used  extensive model-magnitude grids in UV 
and optical filters, for hot WDs (Tlusty) and main sequence and giant stars (Kurucz),  sampling ranges of stellar parameters and reddening \citep{bia24grids}. Owing to the broadness of the filters' passband in the NUV archival data, that includes the 2175\AA~ absorption, model-atmosphere spectra were reddened progressively, with different types of extinction curves, and then  synthetic model magnitudes were computed in  GALEX, Swift, UVIT, HST and other optical filters. 
  Figure  \ref{f_models} shows  sequences of WD model  magnitudes  (\Teff=15k to 200kK,   \logg=7) 
scaled to the target distance of 2463~pc, for two  plausible WD radii, 0.02 and 0.03\rsun, and reddened with a few illustrative values of \ebv: 0.45mag using reddening curves with \rv=3.1 (Milky-Way type) or \rv=2.0, to show the effect of the adopted reddening curve on UV fluxes,  and  \ebv=0.55mag (\rv=3.1) and 0.50mag (\rv=2.0) to show two examples compatible with the observed GALEX-Swift NUV color (vertical line).      A different adopted distance would translate into a corresponding change in radius, for the same magnitude, given that flux scales as R$^2$/D$^2$, and a pair of identical WDs of course would be 0.75mag brighter (twice the flux) than a single WD.  
Figure  \ref{f_models} shows that  the measured GALEX and Swift NUV fluxes, unexplained by T2019's
 scenario and by subsequent works, {\it could be  compatible} with
a hot WD, or a WD pair, with reddening as derived by T2019  or
higher.   However,  the NUV data alone are not sufficient to confirm a possible WD, nor to derive its
 \teff, \logg, radius (hence \Lbol), as the solution strongly depends on  \ebv. Additional FUV colors would be needed to constrain the solution, 
but no far-UV (FUV) data existed for this source (the GALEX FUV detector was off during the exposures, and only NUV images were acquired).

\subsubsection{Astrosat/UVIT FUV observations and Measurements. Results} 
\label{s_uvit}

  Therefore,  we obtained 
Astrosat/UVIT imaging in two FUV filters, the very broad FUV.CaF2 (F148W, \wleff =1481\AA, $\Delta\lambda$=500\AA) 
and the medium-width FUV.Sapphire (F169M, \wleff =1608\AA, $\Delta\lambda$=290\AA). Two UVIT proposals were approved (A09\_058 and A10\_093, P.I. Bianchi). An exceptionally large allocation of observing time was granted to repeat deep exposures sampling orbital phases,  but only part of the approved observations were eventually executed, due to a number of scheduling issues and delays.  The observations are listed in Table \ref{t_uvit}. 
The UVIT data were reduced with the s/w package of \citet{Postmauvit}, that reconstructs the images from the photon-counting data and precisely estimates the effective exposure time accounting for dead-time corrections during the exposures.    The source is not detectable in any of the images.  Based on sky counts at the location of the source (using a 4~pxl radius aperture, sky radii 8-10~pxls and 100 randomly placed apertures in the vicinity of the source),  we obtained 2$\sigma$ upper limits of
$>$25.8~ABmag for FUV.CaF2 ($\sim$2.3$\times$10$^{-18}$\ecmsang at \wleff)  and $>$25.7~ABmag ($\sim$2.2$\times$10$^{-18}$\ecmsang at \wleff) for FUV.Sapphire. 
These limits are  $\approx$4~mag fainter than the GALEX NUV measurements, and of the NUV fluxes in our HST data (Section \ref{s_stis}).

The FUV upper limits  rule out the presence a hot stellar source (see, e.g., FUV-NUV model colors  in \citet{guvmatch20} and \citet{bia24hs}).
Although a hot WD is not present in this case, we show the model predictions (Figure \ref{f_models}) as they may be useful for planning observations in other similar cases.   The  UVIT images, reconstructed from the photon-counting data, are made publicly available online (Section \ref{s_datapublic}).

\subsection{Can a non-degenerate stellar companion account for the NUV flux?}
\label{s_star}
 A main-sequence stellar companion was  ruled out by T2019's  SED  analysis of the then-available data (albeit with the unaccounted discrepancy in the archival GALEX and Swift NUV broad-band fluxes, as we noted), based on the argument that if a companion of mass~$\sim$3.3\Msun  were a non-degenerate star, its luminosity would be detectable in the SED.   
 An alternative scenario was proposed by  van den Heuvel \& Tauris (2020, VdHT20) who 
 suggested that  a close pair of K0-4V stars (2$\times$0.9\msun) could account for the inferred companion mass and have a combined luminosity compatible with the overall SED presented by T2019. Such scenario would account for the companion's presumed mass, but not for the NUV flux, while our proposed alternatives would (Section \ref{s_wd} and here below).  VdHT20 also suggested a 1\Msun red giant primary 
to be more plausible, given the lack of X-ray emission and the surface chemistry of the RG.  Thompson et al. (2020, T2020) responded with arguments favouring a 3.2$^{+1}_{-1}$\Msun RG primary, considering the luminosity.   
The new parallax from Gaia DR3 places the object significantly closer than Gaia DR2, at about 80\%  the distance used by T2019, implying a downwards revision of the RG's  luminosity and radius.

The  presence and the nature  of a stellar companion that could account for both the inferred mass ratio in the system, and the archival NUV fluxes, was successfully tested, and we believe conclusively proven,  by HST UV$-$optical STIS spectroscopy described below.

\begin{figure}[!ht]
\center{
  \vskip -2 cm
\includegraphics[scale=0.14,angle=0]{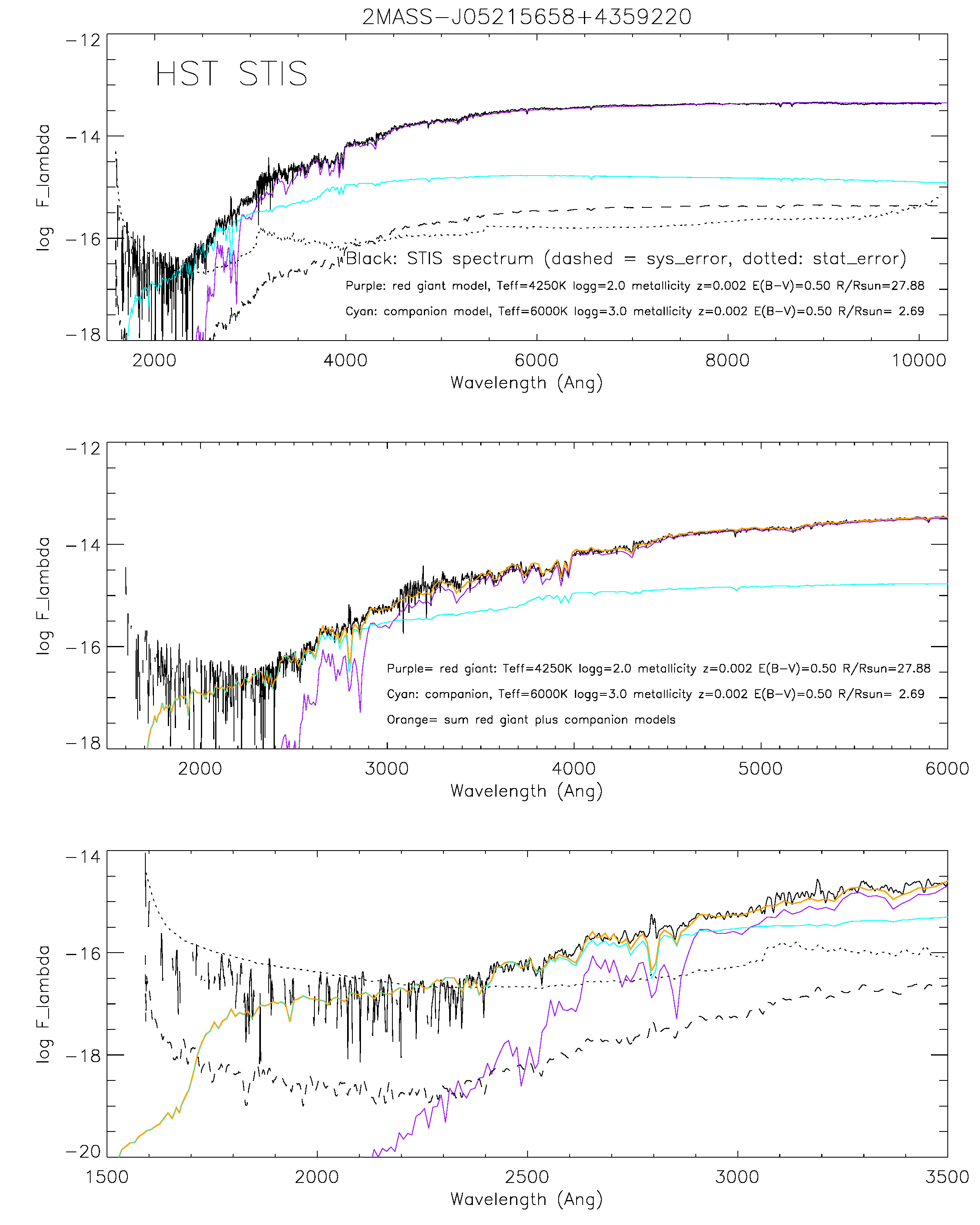}
}
\caption{\footnotesize
{STIS UV+optical spectra (G230L+G430L+G750L combined, flux in \ecmsang) plotted in black, with the best-fit two-component stellar models: \teff=4250~K, \logg=2.0 for the red giant (purple), and \teff=6000K, \logg=3.0 for the companion (cyan). A metallicity of ten times less than solar provides the best match to all major spectral features for the derived \logg, and an extinction of \ebv=0.5 (assuming a Milky-Way type extinction, with \rv=3.1) yields the overall best match to the SED across the whole range. The summed flux of the two models  is shown in orange (middle and bottom panels). With the Gaia DR3 distance, scaling the best-fit-model fluxes to the STIS fluxes accounting for the derived extinction,  yields radii of \rrg=27.8 and \rcomp=2.7\rsun.  In the top and bottom panels the systematic error (dashed line) and statistical error (dotted lines) of the STIS spectra are also shown. Shortwards of $\sim$ 2400\AA~ the data yield only an upper limit to the flux (see Figure \ref{f_upperlimit}). }
\label{f_2stars}
}
\end{figure}

\begin{figure}[!ht]
\center{
\includegraphics[scale=0.20,angle=0]{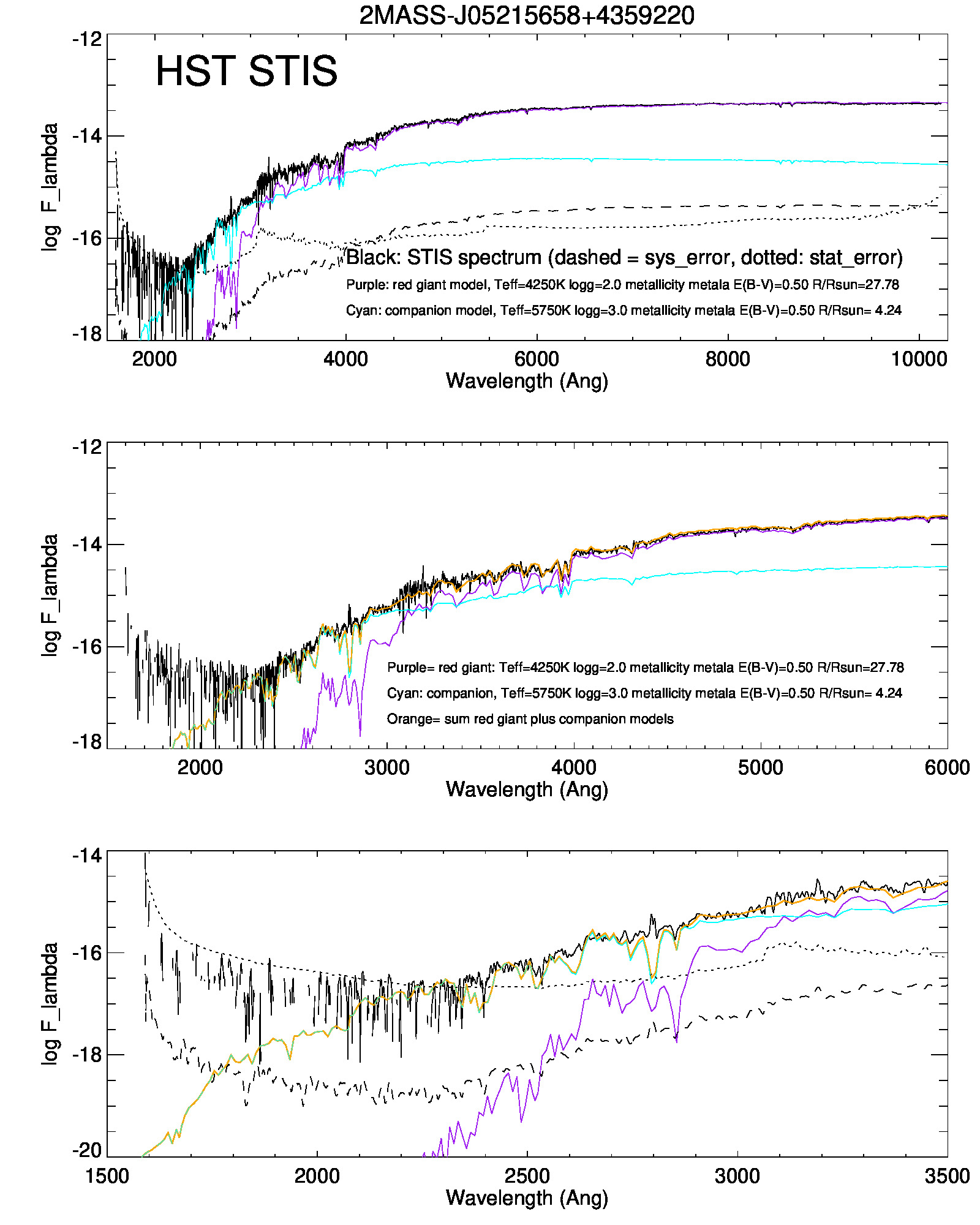}
}
\caption{\footnotesize
  {As in the previous figure: STIS UV+optical spectra (G230L+G430L+G750L combined, flux in \ecmsang) plotted in black, with the best-fit two-component stellar models, using models with metallicity Z=0.006: The red-giant results do not differ from the previous case except for a slightly lower radius: \teff=4250~K, \logg=2.0, \rrg=27.7\Rsun  (purple model), but a worse fit of broad line features is seen between 3000 and 4000\AA, largely compensated by adjusting the companion's parameters:  \teff=5750K, \logg=3.0,  \rcomp=4.2\Rsun (cyan model). Extinction of  \ebv=0.5 still yields  the best solution. The summed flux of the two models  is shown in orange (middle and bottom panels).
    In the top and bottom panels the systematic error (dashed line) and statistical error (dotted lines) of the STIS spectra are also shown. }
\label{f_2starsz0006}
}
\end{figure}

\begin{figure}[!ht]
\center{
\includegraphics[scale=0.128,angle=0]{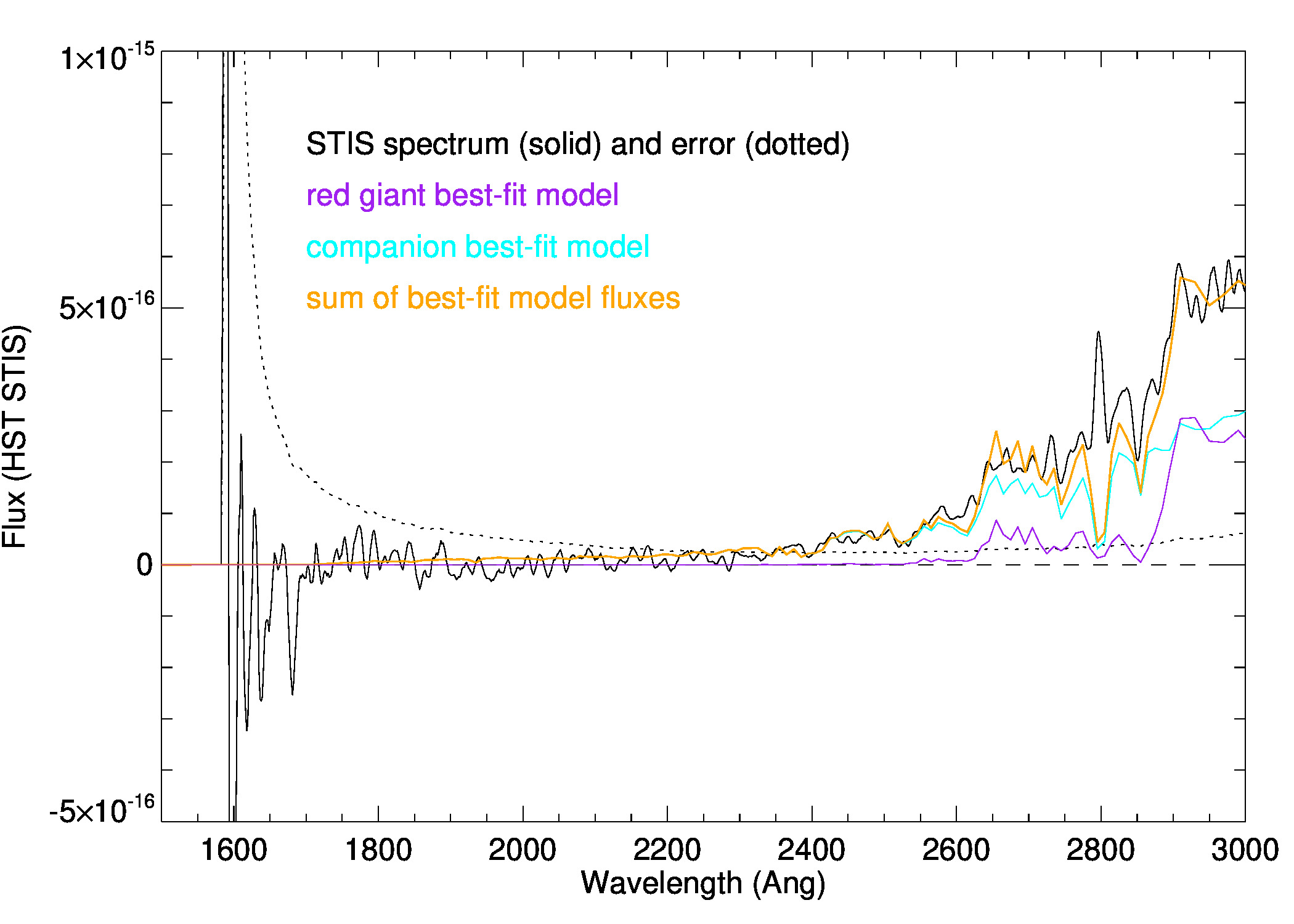}
\includegraphics[scale=0.110,angle=0]{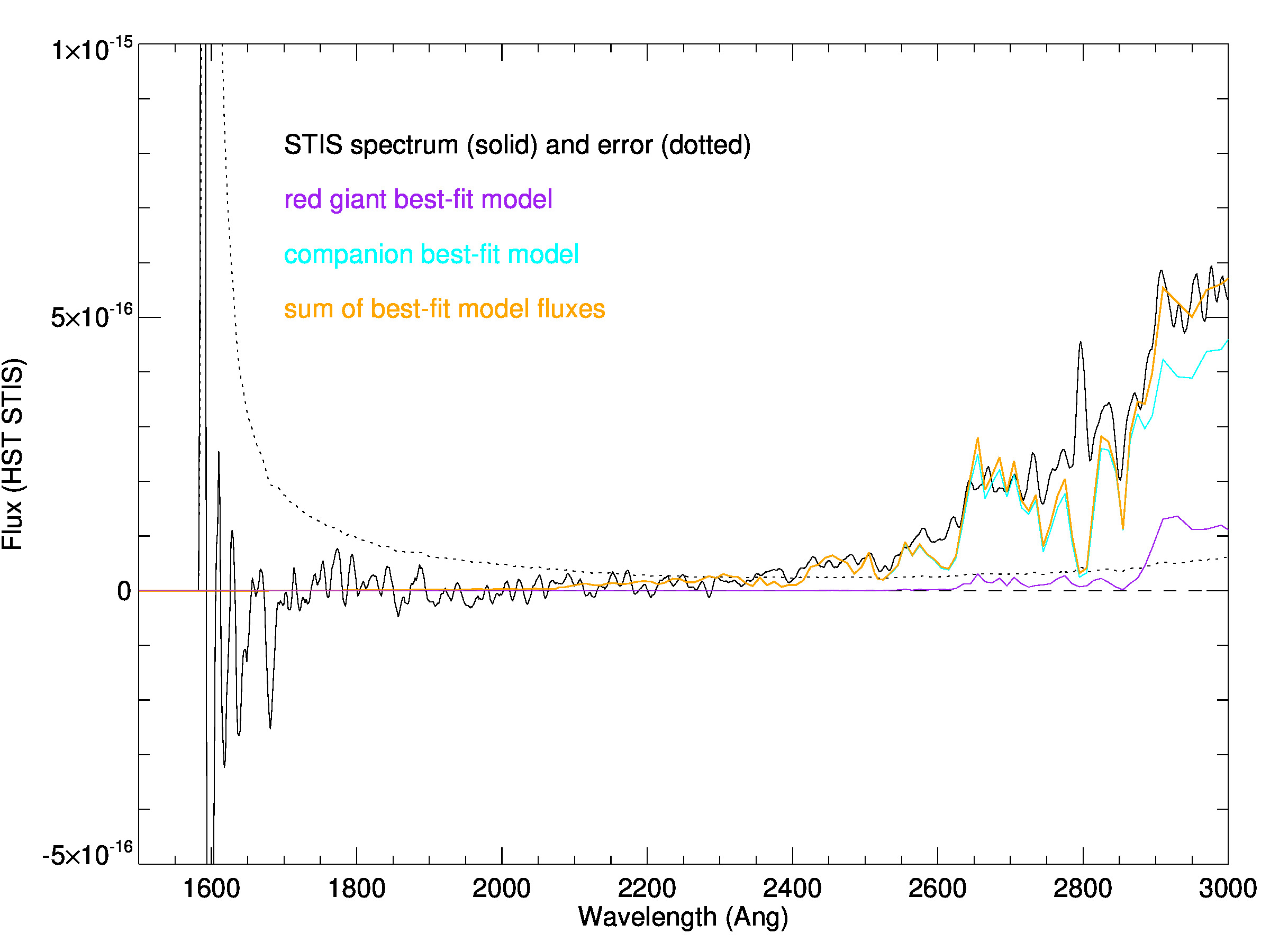}
}
\caption{\footnotesize
The shortest wavelength portion of the STIS spectrum, here with flux (\ecmsang) plotted in linear scale to better distinguish significant flux from noise level, and to illustrate the sensitivity of this range to the companion, whose  flux becomes stronger than the red giant's flux. 
Without this spectral region, the rest of the spectrum could be fit by a wider combination of \Teff and \Ebv values for the stellar pair.  Models with metallicity 
Z=0.002 are used in the top panel, and Z=0.006 
in the bottom panel, as in Figures \ref{f_2stars} and \ref{f_2starsz0006} respectively.  The different metallicity hardly affects the red-giant continuum flux over a wide optical-IR range, as shown in the two previous figures, but the difference becomes significant in this range, where the fit to the composite spectrum is compensated by the companion's solution, requiring a  radius of 4.2\Rsun in the lower plot, $vs$ 2.7\Rsun in the upper plot, and a slightly cooler \teff. 
\label{f_2starsuv}
}
\end{figure}

\begin{figure}[h]
\center{
\includegraphics[scale=0.55,angle=0]{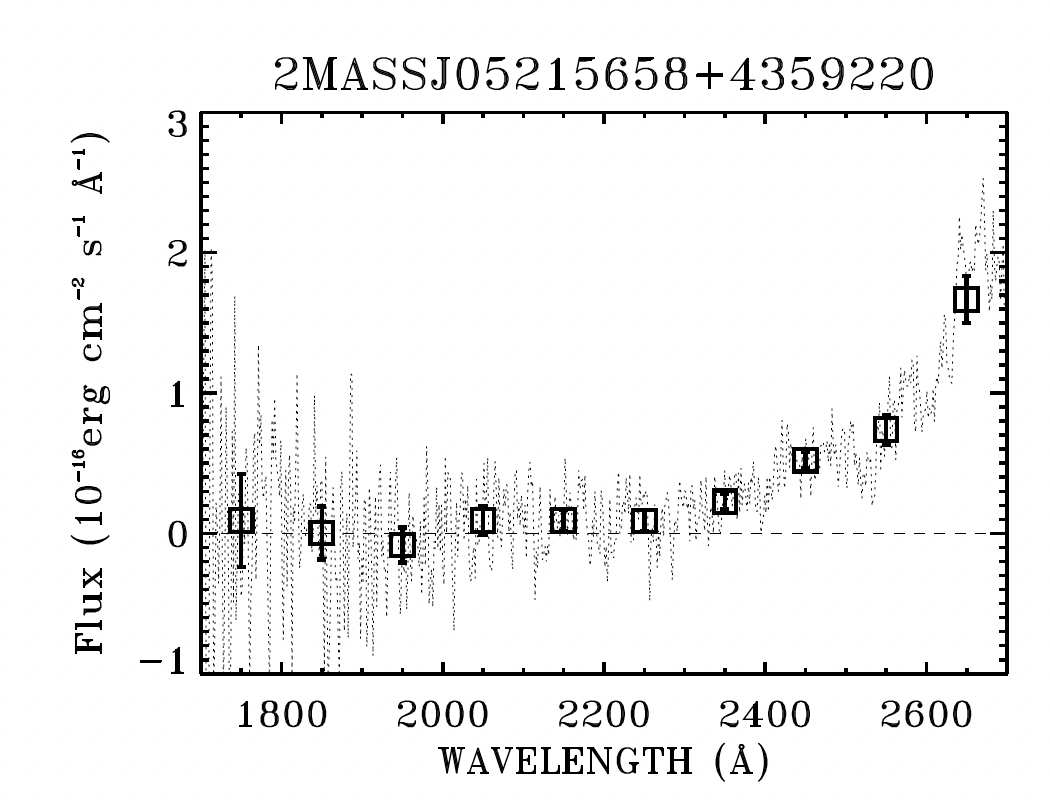}  
}
\caption{\footnotesize
 The STIS G230L combined datasets oemq01010+oemq01050
 (dotted line) are consistent with an upper limit of 
10$^{-17}$\ecmsang~ 
shortwards of $\sim$2300\AA. The squares are averages in 100\AA~ bins. 
\label{f_upperlimit}
}
\end{figure}

\subsubsection{HST STIS Spectroscopy}
\label{s_stis}

We obtained HST/STIS spectra covering the very wide range from $\sim$1,600\AA~ to 10,230\AA, by using three STIS gratings: G230L, G430L and G750L (dispersion = 1.58, 2.73 and 4.92\AA/pxl respectively). Details of the observations are given in Table  \ref{t_hst}.  The main objective was to obtain a UV spectrum, given that information in this wavelength range  was critically missing from all previous studies. The addition of the optical spectrum with the  G430L and  G750L gratings provided simultaneous coverage at longer wavelengths, enabling a robust and consistent modeling of the entire binary system.  Combining UV and optical spectra taken at different times would make the results inconclusive, given the known flux variability of the red giant in V-band, both with  orbital period and long-term variations.    In the original proposal, we had envisaged using the STIS grism to cover at once also the FUV range; however, after the HST proposal was approved and before Phase~2 was finalized, our UVIT observations had began (Section \ref{s_uvit}), showing no FUV fluxes at the expected level in the hot-WD companion hypothesis (Section \ref{s_wd}). Therefore, we switched the configuration to the G230L grating, which yields much better S/N in the range where flux from the source was expected, and was indeed detected (Figures \ref{f_2stars} to \ref{f_2starsuv}).  The STIS data reduction followed the procedure  used for the  CALSPEC database\footnote{http://www.stsci.edu/hst/instrumentation/reference-data-for-calibration-and-tools/astronomical-catalogs/calspec}.  The extracted combined STIS spectrum is also made available in electronic form  (Section \ref{s_datapublic}).

\subsubsection{Modeling the STIS spectra. Parameters of the Stellar Pair.}
\label{s_2stars}

The STIS spectrum, shown in Figures \ref{f_2stars} to \ref{f_2starsuv}, was analyzed with Kurucz models.\footnote{We used the grid of Kurucz models by \citet{castalfa}. 
}  To fit the whole spectrum, which is the sum of two components, each one depending on several parameters, and both affected by reddening whose effect is significantly wavelength dependent because the spectral range spans from UV to IR, we anchored the model fluxes to average continuum flux values computed in small regions of the spectrum rather free of lines. We  started with regions in the longest wavelength part of the spectrum where the companion flux is negligible  (about 0.6\%  or 6\% of the red-giant flux, according to our decomposition in Figures \ref{f_2stars} and \ref{f_2starsz0006},  metallicity Z=0.002 and Z=0.006 respectively); the flux of each model with a given  \Teff (and \logg, metallicity)  was anchored to the red-giant observed flux in a long-wavelength continuum region,
and \Ebv was varied - for each model - in an ample range with small steps; a value was constrained from the slope that best matches the overall shape of the spectrum at long wavelengths and does not exceed the observed fluxes where the companion's flux becomes relevant. A companion's contribution to the total flux was eventually estimated to be $\lesssim$18\% or 32\%  (for the two metallicity values shown) shortwards of $\sim$4,000\AA,  and overwhelming the red-giant flux shortwards of $\sim$3,000\AA, as can be seen in Figures \ref{f_2stars} to  \ref{f_2starsuv}. 
The relative contributions in these regions depend on the model parameters of both stars, and the best match across the whole wavelength range  is achieved with iterations.  The comparison of broad absorption bands in the respective models give a qualitative indication of metallicity and gravity, as shown by the differences between Figures \ref{f_2stars}  {\it vs} \ref{f_2starsz0006},  and in Figure \ref{f_2starsuv}, top {\it vs} bottom panel. 
The solution could be further constrained if metallicity and gravity were precisely derived from high-resolution, high-S/N line analysis spectra, that should be possible at least for the red giant \citep{elbadry24}.

Adopting the Gaia DR3 distance of $D$=2463pc, the best overall fit of both continuum and main absorption bands across the entire range of the  combined STIS spectra (1,600$-$10,230\AA) is obtained by the sum of two stars: a red giant with \Teff=4,250K, \logg=2.0, \rrg=27.9\rsun, and a companion with \Teff=6,000K, \logg=3.0, \rcomp=2.7~\rsun, using models with metallicity of one tenth the solar value, that better match the main absorption features in the UV and blue range (see bottom panels in Figures \ref{f_2stars} and \ref{f_2starsz0006}, and  Figure \ref{f_2starsuv}). Using models with metallicity Z=0.006, the closest value in the model grid to the metallicity  inferred by T2019 and by others from line analysis in ground-based spectra, the red giant parameters hardly change (\rrg=27.78\rsun),  but  parameters of the companion require \teff=5,750K and \rcomp=4.2\Rsun to match the UV region, where the red-giant flux varies the most with metallicity (Figure \ref{f_2starsuv}).  
The values of \logg~  were initially fixed from the preliminary line analysis by 
El-Badry (priv. comm.) of high-resolution, high S/N  Keck-spectra, since those data are more sensitive to this parameter; however, we then varied also  \logg ~ values and could not obtain an equally good SED fit across the whole spectral range by adjusting other parameters.  Extinction was constrained at \ebv=0.50$\pm$0.02mag; small changes would cause large discrepancies in the shortest wavelengths part of the spectrum, or inconsistencies across the entire available range, that could not be compensated by variations of \Teff or other parameters.  

The  STIS spectral coverage extending to UV wavelengths allowed us to refine the stellar parameters of the hotter companion better than is possible from other studies, while the red-giant parameters obtained from the Keck-spectra line analysis (El-Badry, priv. comm.) 
were essentially confirmed. The comparison of two different solutions in Figure \ref{f_2starsuv} illustrates the unique leverage of the UV spectral portion for achieving a consistent solution of all parameters, including stellar radii, as well as the need to constrain metallicity and gravity from high-resolution line analysis, in order to further restrict the acceptable combinations of parameters.  

 The extracted STIS spectrum, convolved with the GALEX NUV transmission curve, yields a synthetic GALEX-NUV magnitude of 21.45~ABmag (19.77 in the Vegamag system). The value is in perfect agreement with the GALEX archival measurements, in spite of the latter having large uncertainties because the source is included only in GALEX  images with very short exposures (Table \ref{t_swiftgalexdata}).

  \subsection{Chromospheric Line Emission from the Red Giant}
\label{s_chromospheric}

\begin{figure}[!ht]
\center{
\includegraphics[scale=0.25,angle=0]{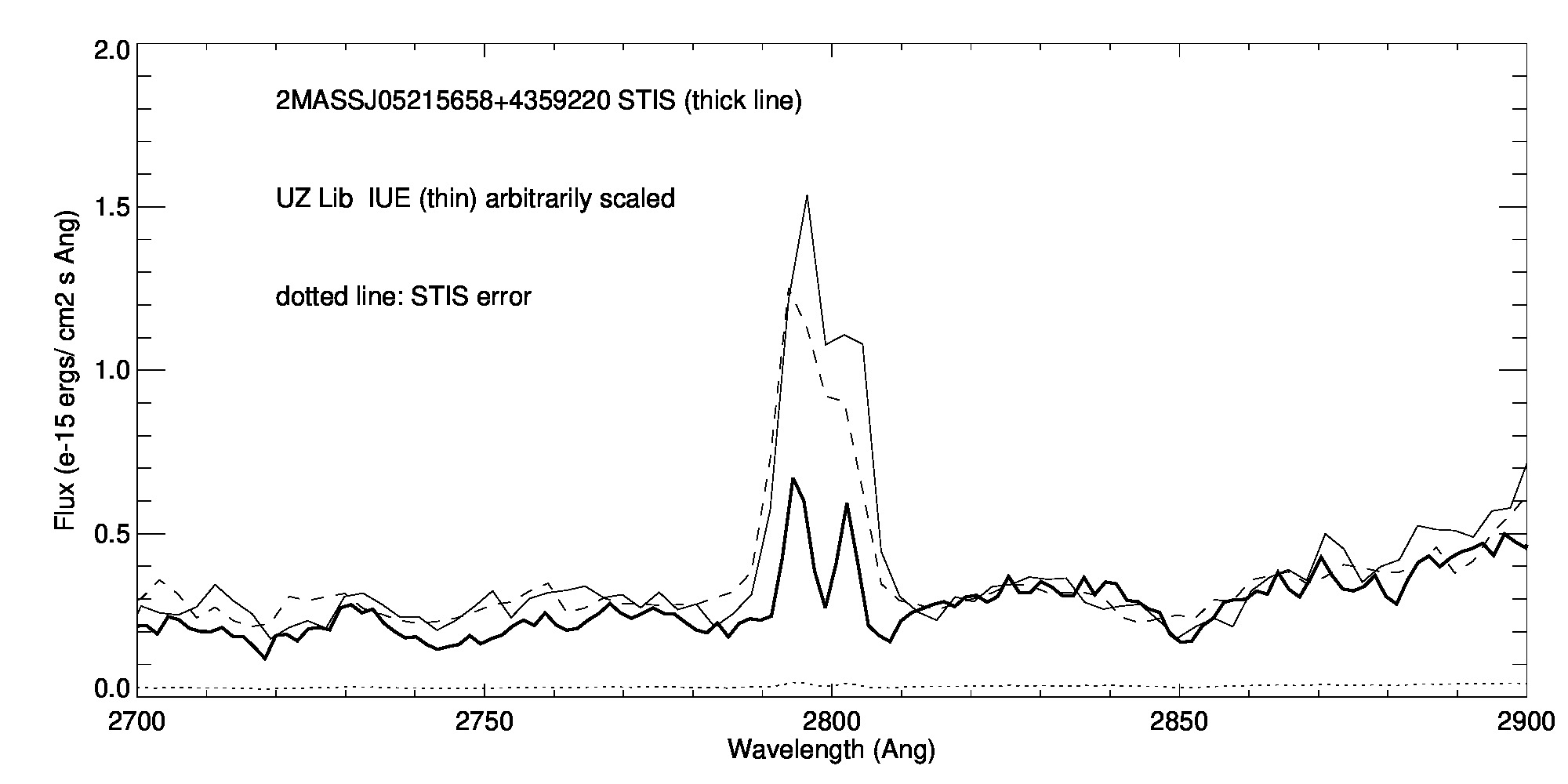}
}
\caption{\footnotesize
 MgII doublet emission in the STIS-G230L spectrum of 2MASS-J05215658+4359220 (thick line) compared with two IUE spectra of UZ~Lib, a chromospherically active red giant (thin solid and dashed lines). The fluxes of each IUE spectrum (LWP5711 and LWP5764) have been scaled  to match the local continuum level of our target around the MgII line (LWP5711 and LWP5764). 
 A caveat must be considered in the comparison: 
in 2MASS-J05215658+4359220 the companion significantly contributes to the flux at these wavelengths (Figures \ref{f_2stars} and \ref{f_2starsz0006} bottom panel, and Figure \ref{f_2starsuv}); subtracting its contribution (cyan models in Figure \ref{f_2starsuv}) would make the MgII emission much stronger.  A simple scaling by radius and distance is not possible, because differential reddening should also be accounted for, which is uncertain for UZ~Lib. 
 Although sampled with a $\sim$2.66\AA~ step, the IUE spectra have an actual  resolution of $\sim$6\AA, lower than the STIS G230L spectra. The STIS resolution has not been degraded in the plot. 
\label{f_mgii} 
}
\end{figure}

 When planning the HST spectroscopy, we had also estimated whether chromospheric and transition zone (TZ) line emission from an active red giant could possibly be sufficient to account for the GALEX and Swift NUV broad-band flux. 
 A non-detection of UV continuum flux, ruling out a stellar companion(s) or an accretion disk,  would have favoured the most exciting among the possible scenarios: it would have confirmed a non-interacting  BH as the companion to the red giant. We had gauged the likelihood of this scenario by scaling nearby  ``analogs'' of the RG primary with existing archival spectra, and found it unlikely to account for the measured NUV fluxes. 
The STIS spectral exposure proved to be adequate to test our different postulated scenarios; the spectra, analyzed in the previous section, show with good S/N both the continuum flux and absorption lines of a stellar companion hotter than the red giant, as well as chromospheric emission from the red giant.
The MgII doublet at $\lambda$$\sim$2800\AA, seen in emission in the STIS spectra, is shown in Figures \ref{f_2starsuv}  and  \ref{f_mgii}. Measuring the emission flux above the local continuum would be misleading, given the  presence of the same transitions in absorption in both the red giant and the companion spectrum (see Figures \ref{f_2stars} and \ref{f_2starsuv}, purple and cyan models), and the current uncertainty of the  model fits in metallicity and gravity would propagate on the flux measurement. Instead, we compared the MgII emission seen in the STIS spectrum with that of a possible analog, sufficiently near that UV archival spectra are available.   
UZ~Lib (K0III) is an active  RS CVn variable, possibly of the  FK~Comae class  (e.g., Strassmaier 2009), 
with some stellar parameters similar to 2MASSJ05215658+4359220; it  also displays photometric variations  of similar amplitude, 
  although it  has a shorter orbital period. 
For UZ~Lib two archival IUE spectra in the NUV range exist; they show strong MgII$\lambda$2800 emission. Some apparent continuum flux at the long-wavelength end of the IUE spectral range is likely red leak. 
Given the difference in reddening and distance between the two systems, in resolution between IUE low-resolution ($\sim$6\AA ) and STIS G230L, and the additional caveat that the NUV  flux of 2MASSJ05215658+4359220 has a major contribution by the hotter companion, overwhelming the red giant continuum in this range (Figure \ref{f_2starsuv}), we scaled the UZ~Lib IUE spectra such that the local continuum surrounding  the MgII emission matches the STIS flux level of 2MASSJ05215658+4359220. The MgII emission is apparently stronger in UZ~Lib, by a factor of about two (Figure \ref{f_mgii}).  But if the companion's contribution to the flux (cyan models in Figure \ref{f_2starsuv}) were subtracted from the STIS spectrum of 2MASSJ05215658+4359220, the red giant's MgII emission would appear stronger than in UZ~Lib.   

 Assessing whether the analogy of the red-giant chromospheric activity extends to the variability characteristics of UZ~Lib type variables or the FK~Comae class, where spot activity is enhanced in one hemisphere, related to binary interaction of a close or coalescing pair, would require UV observations at different orbital phases to monitor possible variability.   We have limited our HST proposal to a  one-time observation, given that the exposure times across the entire spectral range, but especially the UV,  were estimated based on various scenarios that we had postulated, and calculated hypothetically, and it would not have been worth risking more observing time until the exposure times were proven appropriate by the first observations presented here. 

\subsection{Morphology. HST WFC3 imaging data} 
\label{s_wfc3}

\begin{figure}[!ht]
\center{
\includegraphics[scale=0.65,angle=0]{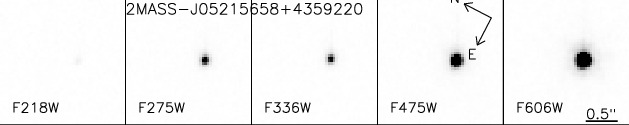}
\vskip 0.1cm
\includegraphics[scale=0.65,angle=0]{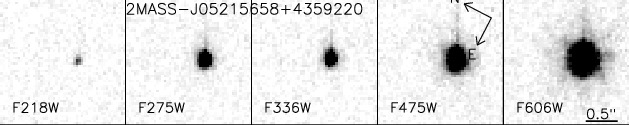}
\vskip 0.1cm
\includegraphics[scale=0.65,angle=0]{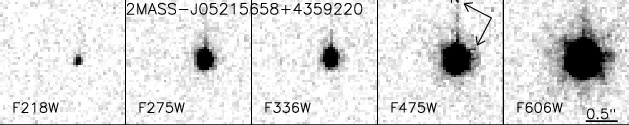}
}
\caption{WFC3 images of the source in the five filters arranged from shorter to longer effective wavelength; repeated in different rows using increasing contrast. Images are 50x50pxl cutouts (one pxl is 0.04\as). For the dithered exposures, the pictures were produced from the drizzled images, therefore the resolution is slightly degraded with respect to the full resolution of individual exposures.
\label{f_wfc3} }
\end{figure}

HST/WFC3 imaging was acquired in filters F218W, F275W, F336W, F475W, and F606W.
The filter choice and exposures were tuned to optimize derivation of physical parameters and extinction for a range of possible stellar companions, based on simulations with our model grids.  In particular, two UV filters are indispensable to disentangle \ebv~ and \Teff if the  companion were a hot stellar source (e.g., \citet{bia14m31}). WFC3 does not have FUV filters, thus F218W and F275W were included.  The WFC3 imaging was obtained 
to measure broad-band fluxes at wavelengths extending to NUV down to the faintest limits we could predict, given that the STIS spectroscopic exposures were estimated based on postulated scenarios yet to be tested, including the presence of a stellar companion hotter than the red giant, but whose \Teff (or even presence) was in fact not predictable with previous data.
We only imaged a small area around the target, using the UVIS2-M1K1C subarray option, which significantly reduces buffer dumps, and allowed us to accommodate all exposures  in one orbit, 
at the expense of complicating the photometric measurements due to lack of field stars to individually determine the point-spread-function (PSF) in each image. Exposures were dithered, for mitigation of
hot pixels and cosmetic defects in the array.  Exposures in the red filters were split in very short exposures (using CTE mitigation strategy) so to not saturate the red giant primary.  Details of all HST data are compiled in Table \ref{t_hst}.  Given the good quality of the STIS spectra across the entire range, and the difficulty to obtain precise photometry from the sub-array images, the analysis to derive stellar parameters was performed only on the STIS spectra, which contain information on both continuum fluxes and line features.  A portion of the WFC3 images around the target is shown in Figure \ref{f_wfc3}; they are the highest-resolution imaging available of the source and its surroundings from 218nm to 606nm, providing information on the morphology, and they further confirm that the source of the UV flux is not an unrelated neighbor.

The WFC3 images  in Figure \ref{f_wfc3} are arranged from the shortest to the longest effective wavelength of the filters (left to right panels). The images are repeated in subsequent rows,  with different contrast, to make more details appreciable; in each row we keep the same constrast relative to the background in all filters, to visualize the large  dynamic range in flux across the sampled wavelength spectrum.  Again the lack of other field sources in the images, preventing derivation of image-specific PSF, makes it difficult to examine whether the source is slightly resolved, or not point-like, in F275W or F366W, where its flux is expected not to be completely overwhelmed by the red giant, according to the STIS spectra.    For a quantitative analysis of the source morphology,  custom 4x-oversampled PSFs from the WFC3 PSFSTD libraries for each filter (or the closest filter in effective wavelength to the observed one) were created, as  appropriate for the true detector position taking into consideration the placement of the subarray and source. Then {\it photutils} (python) was used to do PSF-fitting photometry,  a single pass of detection was run to measure objectively whether there was evidence for a semi-resolved pair. There is no evidence in the residual images; for some images the subtraction of the detected source model is nearly perfect, in some cases a residual feature is seen but it is symmetric around the source, suggesting a mismatch of the PSF model rather than an extra component.  
  From this objective analysis, and from visual inspection,  the source appears unresolved. A WFC3 pxl (0.04\as) at the distance of our target projects to $\sim$100A.U. on the sky (0.000477648~pc).   Assuming that some separation or elongation would be detectable at a resolution of $\sim$2.5pxls, i.e. 0.1\as,   an unresolved source implies an upper limit to the orbital separation of {\bf a}$\times$sin({\it ~i~})$\leq$246A.U., or $\leq$1891 red-giant radii (using the value of R$_{RG}$=28\Rsun derived in section \ref{s_stis}), much larger than is estimated for this pair (Section \ref{s_conclusions}). 
In sum, the broad-band photometry was planned in order to acquire a very deep SED measurement in a short time (one orbit only), in case the STIS exposures failed. The lack of close neighbors in the images further confirms the association of the target with the Gaia counterpart.

\subsection{Current Stellar Masses and Evolutionary Implications}
\label{s_evol}

\begin{figure}[!ht] 
\center{
\includegraphics[scale=0.20,angle=0]{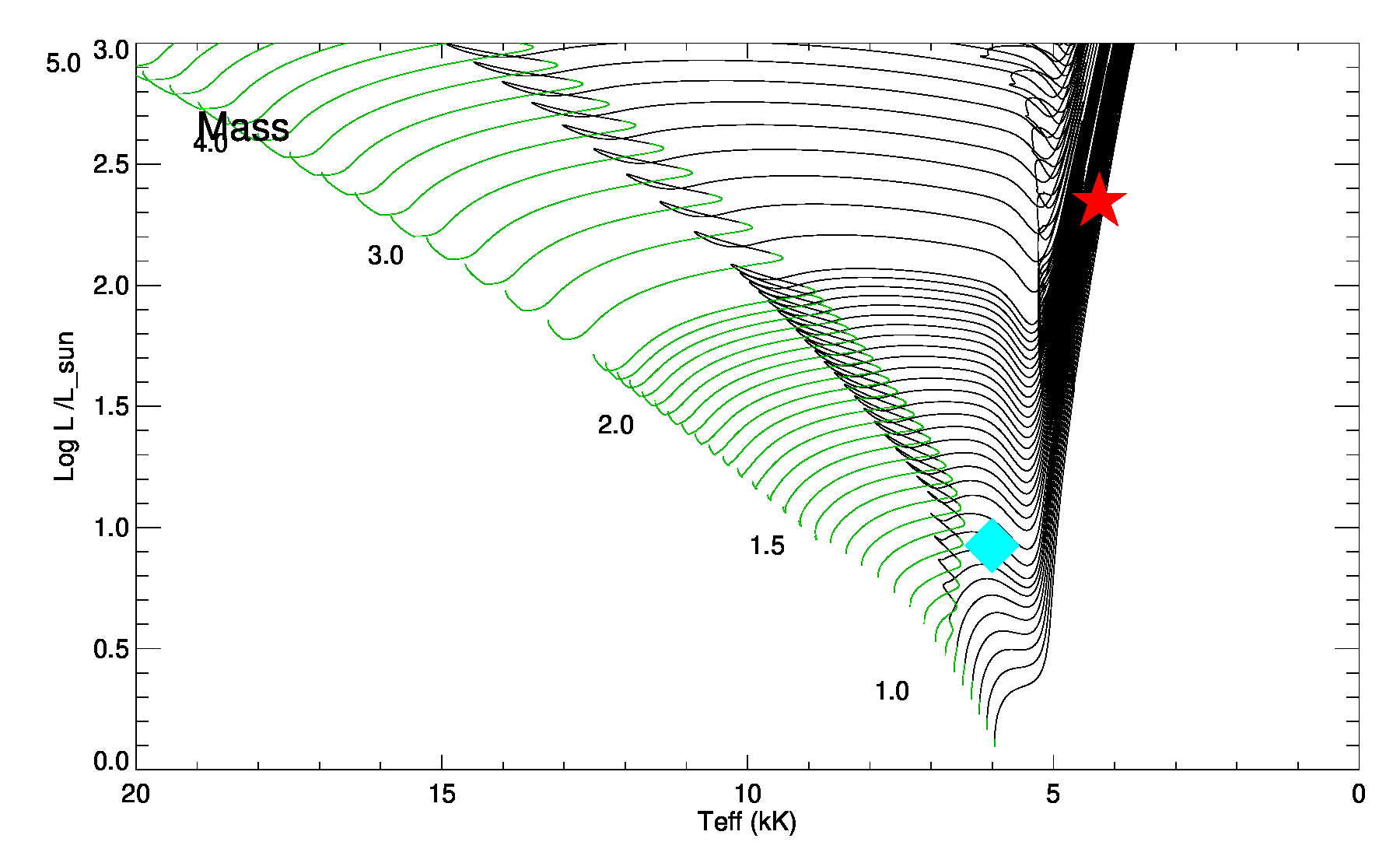}
\includegraphics[scale=0.14,angle=0]{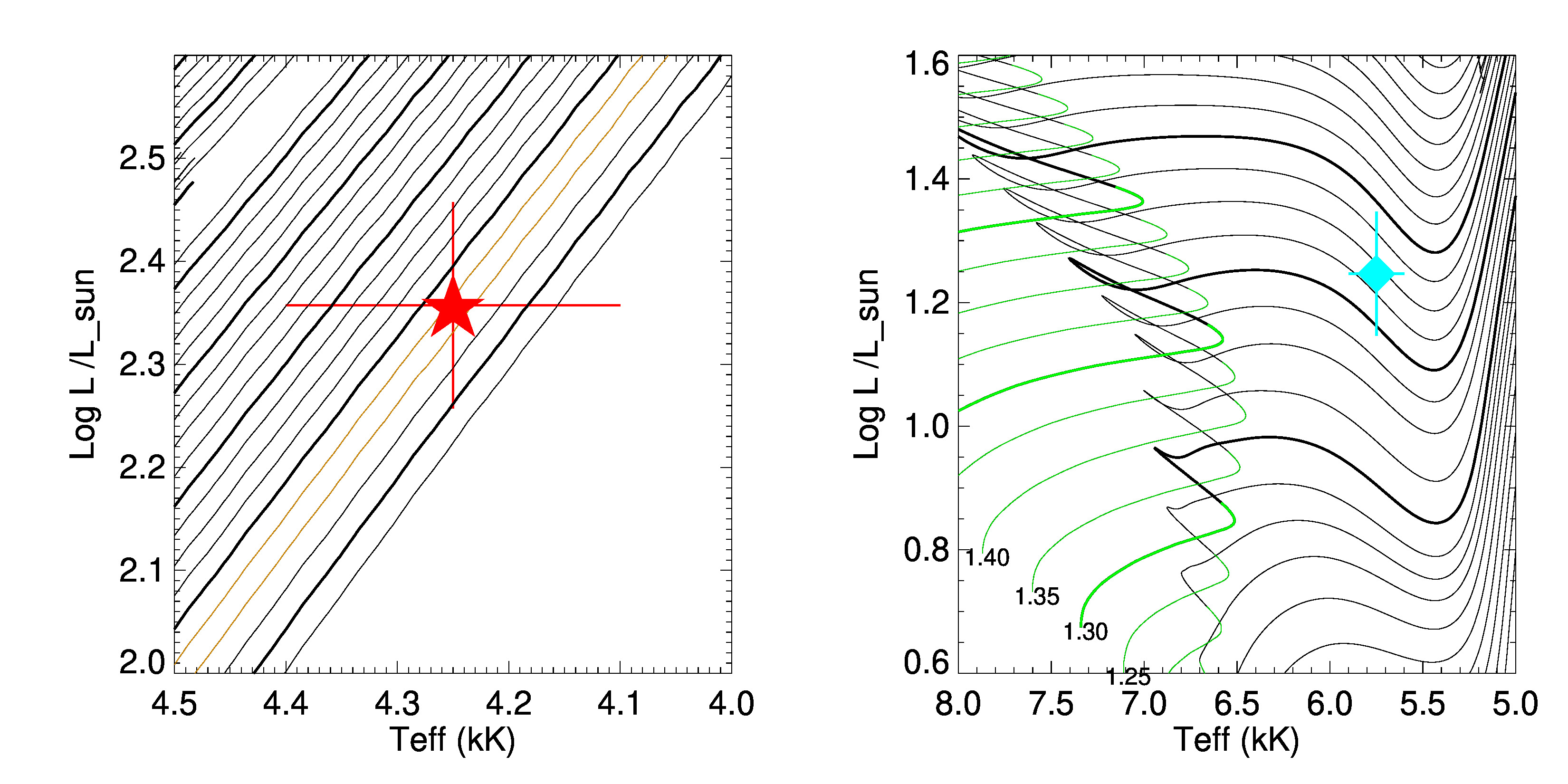}
 }
\caption{HR-diagram of PARSEC evolutionary tracks for metallicity Z=0.006; the main sequence phase is shown in green. The red star marks the red-giant position and the cyan diamond the companion. 
 The bottom panels zoom around the locus of each star: in the left panel (red giant) tracks with masses of 0.9, 1.1, 1.3, 1.5, 1.7\Msun (from bottom-right towards top-left) are marked with thick black lines, and tracks for masses of 1.0 and 1.05\Msun are shown in orange; in the right-bottom panel, thick-line tracks have masses of  1.3, 1.5 and 1.7\msun.  
\label{f_tracks006} }
\end{figure}

\begin{figure}
\center{
\includegraphics[scale=0.14,angle=0]{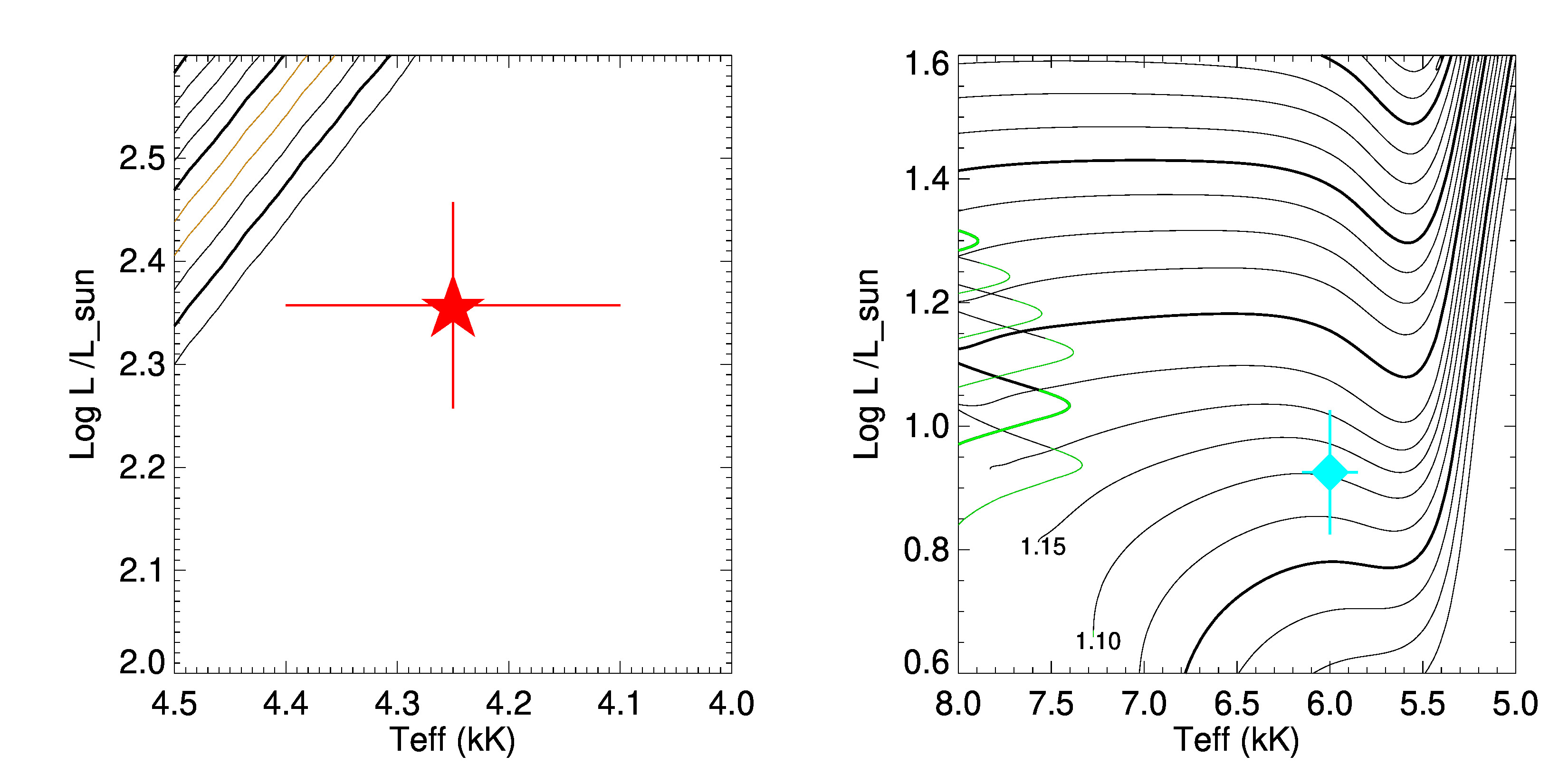}  
 }
\caption{HR-diagram of PARSEC evolutionary tracks for metallicity Z=0.002; the main sequence phase is shown in green. The red star marks the red-giant position and the cyan diamond the companion, using the parameters derived with models of metallicity Z=0.002 (Figure \ref{f_2stars}). 
In the left panel, the loops of the evolutionary tracks  do not reach the \Teff derived for the red giant; in the right panel, thick-line tracks have masses of 1.0, 1.3, and 1.5\msun.  
\label{f_tracks002} }
\end{figure}

 In Figures 
\ref{f_tracks006} and \ref{f_tracks002} the stellar parameters derived in Section \ref{s_2stars} for both the red giant and the subgiant companion are compared with PARSEC evolutionary tracks (\citet{bressan12} and subsequent updates)  for Z=0.006 and Z=0.002. 
The tracks are sampled with fine mass steps in the mass range of the target;  to distinguish them in the complex region of the loops we show them with different colors and thickness, as explained in the figure caption.  The plotted uncertainty in log(\Lbol/\Lsun) is $\approx$0.1 (0.06 from the  \Teff uncertainty and 0.04 from the distance uncertainty, that directly translates into uncertainty in the radius value).   However, as shown in Figures \ref{f_2stars} to \ref{f_2starsuv}, for the lower-metallicity solution a smaller radius and slightly hotter temperature is found for the companion. Therefore, the two  distinct solutions are shown in separate figures. 
The comparison with tracks for Z=0.006 (one third solar)  indicates a red-giant mass around 1\Msun (between 1.35 and 0.85\msun),  and a subgiant companion mass $\approx$1.6$\pm$0.1\msun.   The PARSEC tracks for metallicity of one tenth the solar value do not extend to \Teff values  as low as the red-giant's estimated \teff, but the trend indicates that the inferred mass would then be lower than 1\msun, and the companion's mass $\sim$1.1$\pm$0.1\msun.

\section{Discussion and Summary.}
\label{s_conclusions} 

We investigated the nature of the binary system 2MASSJ05215658+4359220. In particular, this work adds new data in the Ultraviolet range to the existing abundant  data and analyses, and conclusively reveals a red-giant's companion, whose nature was still debated. 
No data existed in the FUV range for the source.  To test the possible presence of a hot compact companion (Section \ref{s_wd}), we obtained Astrosat/UVIT imaging in two filters at far-UV wavelengths. 
To test other scenarios and in particular the presence of a main sequence star hotter but less luminous than the red giant, we obtained HST/STIS spectra and HST/WFC3 high-resolution imaging from the UV to the near-IR range.  

A hot, low-optical-luminosity stellar companion is ruled out by 2~$\sigma$ upper limits obtained in far-UV  imaging with 
Astrosat/UVIT ($\geq$25.8ABmag in FUV.F1 and $\geq$25.7ABmag in FUV.F3) and HST/STIS (flux upper limit of 10$^{-17}$\ecmsang ~shortwards of 2300\AA).

The broad wavelength range HST/STIS spectrum from 1600\AA~ to 10,230\AA~ is well 
matched by the sum of two stellar models: the red giant primary with \Teff=4250K, \logg=2.0, \rrg=27.9\rsun, log(\Lbol/\Lsun)=2.36$\pm$0.1 and a companion with \Teff=6000K, \logg=3.0, \rcomp=2.7~\rsun, log(\Lbol/\Lsun)=0.925$\pm$0.1 for metallicity of one tenth the solar value. 
Extinction was constrained to \ebv=0.50$\pm$0.02mag.  Such combination was found to best match the major absorption features in the UV range (see bottom panel in Figure \ref{f_2stars} and Figure \ref{f_2starsuv}) and the overall flux distribution. Using models with metallicity one third solar, closer to the findings by T2019 and \citet{elbadry24} from ground-based spectra, the whole flux distribution is matched with similar parameters for the red giant (\rrg=27.78\rsun), but slightly different for the companion: \Teff=5750K, \rcomp=4.2\rsun, log(\Lbol/\Lsun)=1.25$\pm$0.1. The continuum SED is still well matched, although some broad absorption bands are less well matched compared to the lower metallicity model (see Figure \ref{f_2starsuv} in particular).
The derived stellar parameters for the red giant are  consistent with the line analysis on high-resolution, high S/N  Keck spectra 
where lines from both stellar components can be resolved (El-Badry, priv. comm). The UV spectra and the analysis of the absolute-calibrated flux SED over the entire STIS spectral range allowed us to refine the companion's parameters and the foreground extinction, which in turn enables derivation of stellar radii from the extinction-corrected fluxes, adopting the Gaia DR3 distance of $D$=2463pc with a 5\% error.     The solution could be further refined if more precise metallicity values and \logg~ will be constrained from high resolution spectra, that should be feasible at least for the red giant.

MgII~$\lambda$2800\AA~ emission from the red giant indicates chromospheric activity  comparable to that of active red giants. In 2MASSJ05215658+4359220, measurement of the line emission must account for the companion's flux contribution that overwhelms the red-giant flux at these wavelengths (Figures \ref{f_2starsuv} and \ref{f_mgii}), making the MgII emission to appear lower than it actually is. 

The STIS spectra, and the UV range in particular,  allowed us to conclusively detect, and to constrain the nature and the stellar parameters of a hotter, lower luminosity companion, simultaneously with the red-giant primary.  However, our HST observations sample only one phase. No orbital-related variability has been explored in UV as yet, while ample literature reports variability in V-mag and radial velocity of the red giant (\citet{elbadry24} and references therein).  The STIS fluxes, convolved with the GALEX transmission curves, give  NUV magnitudes consistent with the measurements in the GALEX archive, but only repeated STIS spectra at selected optical phases can reveal variability (if any) of the hotter component of the system.   \citet{elbadry24}  have also re-analyzed the available photometric data of the target  (V and $g$ band) between 2015 and 2022; the light curves  show a 20\% total amplitude flux variation in V-band up until 2018, then the variability (analyzed by folding the data with the orbital period) seems to have subsided or become more complex in recent years (El-Badry, priv. comm). Similarly, a complex variation over time-scales of years of the optical light curve (folded with the orbital $P$=82.2days)  was illustrated by T2019. According to our spectrum decomposition shown in Figures \ref{f_2stars} and \ref{f_2starsz0006},  the companion flux is  0.6\% or 6.3\% (using models with Z=0.002 or Z=0.006 respectively) of the red giant's flux at the longest wavelengths ($\sim$10,000\AA);   it becomes $\sim$17\% or 32\% of the total flux shortwards of 4000\AA, and it is the predominant source of flux shortwards of $\approx$3,000\AA.   Around 5,000\AA, the companion's flux is 
 $\lesssim$10\% of the total flux, which is essentially the red giant's flux. Therefore, in principle, its contribution is comparable at most to the reported amplitude of the flux variations in V-band. However, the ellipsoidal shape of the light curves, shown by \citet{elbadry24} and by T2019, suggests - as these authors have proposed and demonstrated - that the variability is due to azimuthal asymmetry (and spot activity?) of the red giant's surface in a locked rotation-orbital motion. The possibility of a significant distortion or spot activity  on the red giant surface raises the concern that the spectral decomposition assuming a single-temperature spectrum for the red giant may be an approximation. A more detailed investigation, accounting for possible azimuthal variations of the red giant surface, can be performed on phase-dependent spectra analyzed with updated orbital parameters, and it is not possible with the current single-epoch spectra.  The companion, according to our STIS spectral decomposition, has a much smaller radius than the red giant, and the addition or eclipsing of its flux with orbital phase (assuming an edge-on view) would likely cause a flat-bottom dip in the light curve.     In sum,  a smaller, hotter subgiant companion might be at the other orbital focus from the red giant, but we cannot entirely rule out yet the possibility that the red giant and the subgiant are a very close pair, that underwent recent mass exchange
and might be on their way to coalescing,  both locked together and orbiting a third body.   Repeating the STIS spectra at sample orbital phases would show whether the UV-emitting source is moving in phase, or rather anti-phase, with the red-giant orbit, or moving with it around a third mass.  While investigating variations of the UV flux with Hubble is easily doable, radial velocity measurements of the companion would require a much larger telescope, and high resolution spectra in the blue range. 

 The masses from our spectral decomposition, M$_{RG}$$\sim$1\Msun and M$_{comp}$$\sim$1.6\msun, combined with the orbital period of $P$$_{orb}$=82.2days, imply an orbital  separation {\bf a}=0.51A.U. or  $\sim$110\rsun, i.e. about 4$\times$\rrg.  For the solution yielding M$_{comp}$$\sim$1.1\msun, {\bf a}=0.48A.U., i.e.  $\sim$103\rsun, 3.7$\times$\rrg.  The radius of the red giant's orbit would be {\bf a}$_{RG}$=0.32A.U., 2.4$\times$\rrg, from our derived mass ratio, fairly consistent with the value derived from $P$ and RV amplitude, {\bf a}$_{RG}$=0.34A.U., and  implying, as found in previous works, that currently the red giant is not filling its Roche lobe. Such a small separation also favours a distortion of the red giant surface, and may explain the temperature and luminosity difference between the hemisphere facing the companion and the opposite one, as seen by T12019 and \citet{elbadry24}.  The subgiant companion's RV amplitude should then be $\sim$27.8\kms (for mass ratio 1/1.6), or 40.5\kms for mass ratio 1./1.1.    If both stars were instead orbiting around a third mass, i.e. if we suppose that the red giant and the subgiant are a  coalescing pair  orbiting with RV amplitude of 44.5\kms around the postulated BH,  the orbit radius of the red giant {\bf a}$_{RG}$=0.34A.U.  (from the published $P$ and RV values)  would imply a mass of 2.8\Msun  at the other focus, and a total separation of 0.65A.U.   Such scenario would explain the estimated mass function that suggests a mass lower limit for the undetected companion of 2\Msun (T2019, and El-Badry priv. comm.), larger than what our spectroscopic analysis suggests for the subgiant.

The main result from this work is the definitive detection of a stellar companion, a subgiant hotter and slightly more massive than the red giant primary, implying recent mass transfer from the red giant,  and removing the need (but not the possibility) to postulate a BH companion for  one of the rare standing candidate non-interactive BH binary systems. The simultaneous space-based (abolute-flux calibrated) spectroscopy from $\sim$1,600 to 10,230\AA~ uniquely enabled derivation of stellar parameters for the two stars, and in particular for the hotter companion which is elusive at optical wavelengths where fluxes are dominated by the red giant. However, confirmation that the pair is only comprised of these two components, and that the NUV-emitting source is not a close (possibly merging) companion to the red giant, with  a third mass on a wider orbit, could only come from phase-monitoring of the UV fluxes, and modeling of high-resolution line spectra in the blue range. 
Our initially postulated additional alternative that the NUV flux may arise from an accretion disk, unilkely given the lack of X-ray emission \citep{vandenHeuvelTauris2020}, seems further dismissed by the lack of FUV flux and the presence of absorption bands in the STIS NUV spectra (Figure \ref{f_2starsuv}),  by comparison with common accretion disk models and  studies (e.g., \citet{puebla07,HubenyLong21} and references therein). 
These results also make this source a valuable benchmark for testing binary stellar evolution theories, given the unusual abundance of data and precise measurements, augmenting the scant existing well-studied sample of stripped red-giant binaries (e.g., \citet{elbadry22}).

The current results indicate that the red giant is probably not filling its Roche lobe, 
 although optical-light variability locked with orbital phase (T2019 and references therein) are likely a relic of recent interaction, as is also implied from the mass ratio and from the chromospheric emission shown in this work.  Precise abundance constraints from high-resolution line analysis would critically reduce the uncertainty in the current mass estimates.

\section{Data Publicly Available}
\label{s_datapublic}

Please reference this publication if using the data below. 

\begin{itemize}
\item
The STIS extracted spectra, combined for the three gratings, can be downloaded from: \\
{\small
\url{http:/dolomiti.pha.jhu.edu/uvsky/RGbinary\_2MASSJ05215658+4359220/STISspectra/STIS\_2mj05215.mrg} }\\
The file includes the three-ranges spectra extracted and combined: G230L+G430L+G750L \\
Ascii file. Columns are: WAVELENGTH (\AA),   COUNT-RATE (\mbox{counts~cm$^{-2}$~s$^{-1}$~pixel$^{-1}$}),    FLUX (\ecmsang),     STAT-ERROR  (\ecmsang),   SYS-ERROR  (\ecmsang),  NPTS (the number of co-added spectra for that pixel),   TIME (total exposure time at that pixel in seconds), 
  QUAL (
1=good data, 
0=possibly poor data quality).\\
The DOI: 10.17909/tvvw-4e47 refers to the original HST data products in MAST, not to the re-extracted spectra available here. 
\item
 The UVIT images, reconstructed from the photon-counting archival data, calibrated accounting for dead-time corrections, can be downloaded at:\\
{\small
\url{http:/dolomiti.pha.jhu.edu/uvsky/RGbinary\_2MASSJ05215658+4359220/UVITimages/UVITimages.tar}\\ }
The tar file contains the following .fits files of the reconstructed images, where the ``NORM\_EXPARRAY'' files are the normalized relative response maps across the field, expressed as 0 to 1.  So, the effective exposure time at a pixel position is RDCDTIME multiplied by this image.  The ''NORMEXP\_IMAGE'' is the science image, that must be divided by the product of RDCDTIME and the NORM\_EXPARRAY image on a pixel-by-pixel basis, to obtain counts s$^{-1}$~pixel$^{-1}$  \\
{\footnotesize
2MASSJ05215658+4359220\_FUV\_CaF2\_\_\_MASTER\_NORM\_EXPARRAY\_A10\_093T01\_9000004128.fits\\
2MASSJ05215658+4359220\_FUV\_CaF2\_\_\_MASTER\_NORM\_EXPARRAY\_A10\_093T01\_9000004266.fits\\
2MASSJ05215658+4359220\_FUV\_CaF2\_\_\_MASTER\_NORM\_EXPARRAY\_Merged.fits\\
2MASSJ05215658+4359220\_FUV\_CaF2\_\_\_MASTER\_NORMEXP\_IMAGE\_A10\_093T01\_9000004128.fits\\
2MASSJ05215658+4359220\_FUV\_CaF2\_\_\_MASTER\_NORMEXP\_IMAGE\_A10\_093T01\_9000004266.fits\\
2MASSJ05215658+4359220\_FUV\_CaF2\_\_\_MASTER\_NORMEXP\_IMAGE\_Merged.fits\\
2MASSJ05215658+4359220\_FUV\_Sapphire\_\_\_MASTER\_NORM\_EXPARRAY\_A10\_093T01\_9000004128.fits\\
2MASSJ05215658+4359220\_FUV\_Sapphire\_\_\_MASTER\_NORM\_EXPARRAY\_A10\_093T01\_9000004266.fits\\
2MASSJ05215658+4359220\_FUV\_Sapphire\_\_\_MASTER\_NORM\_EXPARRAY\_Merged.fits\\
2MASSJ05215658+4359220\_FUV\_Sapphire\_\_\_MASTER\_NORMEXP\_IMAGE\_A10\_093T01\_9000004128.fits\\
2MASSJ05215658+4359220\_FUV\_Sapphire\_\_\_MASTER\_NORMEXP\_IMAGE\_A10\_093T01\_9000004266.fits\\
2MASSJ05215658+4359220\_FUV\_Sapphire\_\_\_MASTER\_NORMEXP\_IMAGE\_Merged.fits\\
}

\end{itemize}

\clearpage 

\begin{deluxetable}{clrrlrrccccccccc}
\tabletypesize{\tiny}
\rotate
\tablecaption{Archival NUV  measurements  for 2MASSJ05215658+4359220\tablenotemark{a}   \label{t_swiftgalexdata} }
\vskip -2.cm 
\tablehead{
\colhead{Instrument and Filter}       &  \colhead{Obs.ID } &  \colhead{Ra,Dec center}  &    \colhead{Start Date} &  \colhead{End Date}  & \colhead{exp.time}
&\colhead{      mag    (error)}  & \\
\colhead{}                            &  \colhead{      }             &  \colhead{      }&  \colhead{        }         &  \colhead{    }             & \colhead{ (sec) } 
& \colhead{ } & } 
\tablewidth{0pt}
\startdata
GALEX  NUV &AIS\_57\_1\_69\tablenotemark{b} & 05:21:56.760 +43:59:22.03 & 2012-01-5   & 2012-01-5 & 96.00 (65.89) & 21.44$\pm$0.59~ABmag  & \\ 
GALEX  NUV & AIS\_57\_1\_75\tablenotemark{b} & 05:21:56.610 +43:59:22.01 & 2012-01-5  & 2012-01-5 & 103.05 (70.67) & 21.49$\pm$0.36~ABmag  & \\
& & & \\
UVOT UVM2\tablenotemark{c}    & 00010442002      &05 22 05.240   +43 59 44.54 &	2018-08-09 08:23:52 &	2018-08-09 10:14:56 & 1435.612  (2)  
& $>$ 20.27~Vegamag   \\
UVOT UVM2\tablenotemark{c}    & 00010442003      &05 22 01.996	+43 59 19.90 &	2018-08-10 19:30:19 &	2018-08-11 10:10:55 & 2791.814	(4)  
&  19.75$\pm$0.21~Vegamag\\
UVOT UVM2\tablenotemark{c}    & 00010442004      &05 21 59.182	+43 59 02.73 &	2018-08-12 23:59:49 &	2018-08-13 01:47:55 & 2101.318	(2)  
&  20.22$\pm$0.21~Vegamag\\
UVOT UVM2\tablenotemark{c}    & 00010442005      &05 22 01.732	+43 57 24.03 &	2018-08-15 01:35:49 &	2018-08-15 06:36:55  & 2068.513	(3)  
&  19.90$\pm$0.19~Vegamag\\
UVOT UVM2\tablenotemark{c}    & 00010442006      &05 21 58.697	+43 58 23.97 &	2018-09-08 00:51:26 &	2018-09-08 10:20:52 & 1999.570	(6)  
&  20.06$\pm$0.22~Vegamag\\
UVOT  U\tablenotemark{c}    & 00010442006       &05 21 58.697	+43 58 23.97 &	2018-09-08 00:58:05 &	2018-09-08 10:22:56 &1920.615	(6)  
&  19.77$\pm$0.18~Vegamag \\
\enddata
\tablenotetext{a}{For a compilation of other data at longer wavlengths see El-Badry et al. 2024} 
\tablenotetext{b}{in the first and second row, respectively: IAU name: GALEXJ052156.7+435922 and GALEXJ052156.6+435922, {\it GALEX objid}=  6372921338785432248 and 6372921345227885710, {\it GALEX photoetxractid}=6372921338784382976 and  6372921345226833920 in the GALEX database}
\tablenotetext{c}{from Thompson et al. (2019); Target name: 	BR0521\_4359}
\end{deluxetable}

\begin{deluxetable}{clrllcc}
\tabletypesize{\tiny}
\rotate
\tablecaption{Astrosat/UVIT observations and resulting  measurements  for 2MASSJ05215658+4359220.   \label{t_uvit} }
\vskip -2.cm 
\tablehead{ 
\colhead{Instrument and Filter}        &  \colhead{Obs.ID\tablenotemark{a} }   &    \colhead{Dataset} &  \colhead{Date}  & \colhead{exp.time\tablenotemark{b}}&\colhead{      mag    (  error)}  & \\
\colhead{}                                         &  \colhead{      }&  \colhead{        }         &  \colhead{ (Mean Helioc.JD)     }             & \colhead{ (sec) } & \colhead{(ABmag) }  } 
\tablewidth{0pt}
\startdata
UVIT    FUV.F1 (CaF2)     & A10\_093T01\_9000004 &  A10\_093T01\_9000004128     & 2459236.3073659 & 4133.383   &$>$25.0\tablenotemark{c}    (2$\sigma$) \\
UVIT    FUV.F1 (CaF2)     & A10\_093T01\_9000004 &  A10\_093T01\_9000004266     & 2459291.7881458 & 4251.031   & \\
UVIT    FUV.F1 (CaF2)     & A10\_093T01\_9000004 &  Merged                      &                 & 8384.415   & $>$25.8\tablenotemark{d} (2$\sigma$) & \\
UVIT    FUV.F3 (Sapphire) & A10\_093T01\_9000004 &  
                                                    A10\_093T01\_9000004128      & 2459236.3219791 & 6561.54 & $>$24.5\tablenotemark{c}    (2$\sigma$)\\
UVIT    FUV.F3 (Sapphire) & A10\_093T01\_9000004 & A10\_093T01\_9000004266      & 2459291.85908   & 6865.869 & \\
UVIT    FUV.F3 (Sapphire) & A10\_093T01\_9000004 &  Merged                      & & 13427.442 & $>$25.7\tablenotemark{d} (2$\sigma$)\\ 
\enddata
\tablenotetext{a}{Bianchi's program A10\_093} 
\tablenotetext{b}{RDCDTIME  (Physical Integration times for images, Reduced for Parity Errors, Cosmic Rays  and Missing Frames)}
\tablenotetext{c}{2$\sigma$ upper limit, obtained by measuring countrates of 100 random apertures in the target position and nearby, for aperture size=3pxl (approximately 3\as). For aperture size of 5~pxls we obtain upper limits about half magnitude brighter. Correction for the fraction of flux in the aperture is applied following \citet{Tandon2020}; no extinction correction is applied. }
\tablenotetext{d}{a 4~pixel radius was used for measurements on merged images}
\end{deluxetable}

\begin{deluxetable}{lcccccccccc}
\tabletypesize{\tiny}
\rotate
\tablecaption{HST observations (Bianchi's program HST-GO-16654)  \label{t_hst} }
\vskip -2.cm 
\tablehead{
\colhead{Dataset}        &  \colhead{Target ID}   &    \colhead{R.A.} &  \colhead{Dec.}  &  \colhead{Start Time }  &  \colhead{End Time} & \colhead{Exp. Time} & 
\colhead{Instr.}     & \colhead{Aperture}  &       \colhead{Filter/} & \colhead{Central $\lambda$}   \\
\colhead{}       &  \colhead{      }&  \colhead{        }         &  \colhead{    }  &  \colhead{    }     &     & \colhead{ (sec) }  &\colhead{    }&\colhead{    } & \colhead{Grating    }       &\colhead{    }    }
\tablewidth{0pt}
\startdata
\multicolumn{5}{c}{HST Imaging} \\
IEMQA1010 & 2MASS-J05215658+4359220 & 80.48579541667 & 43.98943333333 & 2022-04-17T07:33:34.060 & 2022-04-17T07:58:10.583 & 0.96 & WFC3   & UVIS2-C1K1C-SUB & F606W & 5887.7124\\
IEMQA1020 & 2MASS-J05215658+4359220 & 80.48579541667 & 43.98943333333 & 2022-04-17T07:35:20.073 & 2022-04-17T07:59:56.760 & 1.39 & WFC3   & UVIS2-C1K1C-SUB & F475W & 4772.1709\\
IEMQA1030 & 2MASS-J05215658+4359220 & 80.48579541667 & 43.98943333333 & 2022-04-17T07:37:21.067 & 2022-04-17T08:03:44.057 & 20.0 & WFC3   & UVIS2-C1K1C-SUB & F336W & 3354.6558\\
IEMQA1031 & 2MASS-J05215658+4359220 & 80.48579541667 & 43.98943333333 & 2022-04-17T07:37:21.067 & 2022-04-17T07:39:08.057 & 10.0 & WFC3   & UVIS2-C1K1C-SUB & F336W & 3354.6558\\
IEMQA1032 & 2MASS-J05215658+4359220 & 80.48584915326 & 43.98944463119 & 2022-04-17T08:01:57.067 & 2022-04-17T08:03:44.057 & 10.0 & WFC3   & UVIS2-C1K1C-SUB & F336W & 3354.6558\\
IEMQA1040 & 2MASS-J05215658+4359220 & 80.48579541667 & 43.98943333333 & 2022-04-17T07:41:08.073 & 2022-04-17T08:13:22.073 & 528.0 & WFC3   & UVIS2-C1K1C-SUB & F218W & 2223.7244\\
IEMQA1041 & 2MASS-J05215658+4359220 & 80.48579541667 & 43.98943333333 & 2022-04-17T07:41:08.073 & 2022-04-17T07:48:46.073 & 264.0 & WFC3   & UVIS2-C1K1C-SUB & F218W & 2223.7244\\
IEMQA1042 & 2MASS-J05215658+4359220 & 80.48584915326 & 43.98944463119 & 2022-04-17T08:05:44.073 & 2022-04-17T08:13:22.073 & 264.0 & WFC3   & UVIS2-C1K1C-SUB & F218W & 2223.7244\\
IEMQA1050 & 2MASS-J05215658+4359220 & 80.48579541667 & 43.98943333333 & 2022-04-17T07:50:46.090 & 2022-04-17T08:20:49.080 & 460.0 & WFC3   & UVIS2-C1K1C-SUB & F275W & 2703.2976\\
IEMQA1051 & 2MASS-J05215658+4359220 & 80.48579541667 & 43.98943333333 & 2022-04-17T07:50:46.090 & 2022-04-17T07:56:13.080 & 230.0 & WFC3   & UVIS2-C1K1C-SUB & F275W & 2703.2976\\
IEMQA1052 & 2MASS-J05215658+4359220 & 80.48584915326 & 43.98944463119 & 2022-04-17T08:15:22.057 & 2022-04-17T08:20:49.080 & 230.0 & WFC3   & UVIS2-C1K1C-SUB & F275W & 2703.2976\\
OEMQ01MIQ\tablenotemark{a} & 2MASS-J05215658+4359220 & 80.48579541667 & 43.98943333333 & 2022-04-17T04:23:12.010 & 2022-04-17T04:26:13.167 & 3.1 & STIS   & F28X50LP & MIRVIS & 0.0 \\
\multicolumn{5}{c}{HST Spectroscopy} \\
OEMQ01010 & 2MASS-J05215658+4359220 & 80.48579541667 & 43.98943333333 & 2022-04-17T04:31:38.987 & 2022-04-17T04:43:45.987 & 727.0 & STIS   & 52X0.5 & G230L & 2376.0\\
OEMQ01020 & 2MASS-J05215658+4359220 & 80.48579541667 & 43.98943333333 & 2022-04-17T04:46:03.977 & 2022-04-17T04:57:31.977 & 600.0 & STIS   & 52X0.5 & G430L & 4300.0\\
OEMQ01030 & 2MASS-J05215658+4359220 & 80.48579541667 & 43.98943333333 & 2022-04-17T05:03:12.970 & 2022-04-17T05:09:40.970 & 300.0 & STIS   & 52X0.5 & G750L & 7751.0\\
OEMQ01050 & 2MASS-J05215658+4359220 & 80.48579541667 & 43.98943333333 & 2022-04-17T05:57:59.977 & 2022-04-17T06:44:29.977 & 2790.2 & STIS   & 52X0.5 & G230L & 2376.0\\
OEMQ01040\tablenotemark{b} & NONE   & 80.3780550487  & 44.02705481179 & 2022-04-17T05:13:16.970 & 2022-04-17T05:18:38.953 & 240.0 & STIS   & 0.3X0.09 & G750L & 7751.0\\
\enddata
\tablenotetext{a}{Acquisition Image for STIS spectrscopy} 
\tablenotetext{b}{Fringe flat for the G750L. ``NONE'' indicates that there was no external target,  just the internal Tungsten lamp}
\end{deluxetable}

{}
\acknowledgments 
 LB and DT acknowledge  support from grant HST-GO-16654.  LB is very grateful to the UVIT support team and to the HST program scientists for help in optimizing and executing the approved observations and to Joe Postma for reconstructing the UVIT images with his s/w package.  LB is extremely grateful to Kareem El-Badry, who shared results from his high-resolution, high S/N Keck spectra prior to publication, and provided most helpful and generous discussions on the nature of the system. E.B. is supported by NSF Grant No. AST-2307146 and by the Simons Foundation.

{\it Facilities:} \facility{UVIT}, \facility{HST (STIS)}, \facility{HST (WFC3)}, \facility{GALEX},  \facility{MAST},  \facility{Gaia} 

\textbf{ORCID iDs}
Luciana Bianchi https://orcid.org/0000-0001-7746-5461


\clearpage 
\begin{thebibliography}{}
\bibitem[Abadie et al(2010)]{abadie2010}Abadie, K., Abbott, B. P., Abbott, R.,  et al. 2010, Classical and Quantum Gravity, Volume 27, Issue 17, id. 173001 (2010), DOI:   10.1088/0264-9381/27/17/173001  
\bibitem[Abbott et al.(2016)]{abbott2016}Abbott, B. P. ,  Abbott, R.,  Abbott, T. D., et al. 2016, \apj 818, L22 
\bibitem[Abbott et al.(2017a)]{abbott2017a}Abbott, B. P. ,  Abbott, R.,  Abbott, T. D., et al. 2017a, \apj 848, L12
\bibitem[Abbott et al.(2017b)]{abbott2017b}Abbott, B. P. ,  Abbott, R.,  Abbott, T. D.,  et al. 2017b, \apj 848, L13
\bibitem[Abbott et al.(2017c)]{abbott2017c}Abbott, B. P., Abbott, R., Abbott, T. D., et al. 2017, Phys.Rev.Lett. Volume 119, Issue 16, id.161101, doI:  
10.1103/PhysRevLett.119.161101
\bibitem[Antoniadis et al.(2016)]{Antoniadis2016}Antoniadis, J., Tauris, T., Ozel, F. et al. 2016, arXiv:1605.01665, DOI:  
10.48550/arXiv.1605.01665
\bibitem[Bailer-Jones(2011)]{Bailer-Jones2011}Bailer-Jones, C.A.L. 2011, \mnras 411, 435
\bibitem[Banerjee(2018)]{banerjee18}Banerjee, S.\ 2018, \mnras , 481,  5123; DOI:   10.1093/mnras/sty2608
\bibitem[Bianchi(2024a)]{bia24hs}Bianchi, L., 2024a, \apjs, in press
  (DOI:  10.3847/1538-4365/ad6e7c ) ; arXiv:2409.04626
\bibitem[Bianchi(2024b)]{bia24grids}Bianchi, L., 2024b, in preparation
\bibitem[Bianchi et al.(2014)]{bia14m31}Bianchi, L.,  Kang, Y., Hodge, P., et al. 2014,  J. Adv. Space Res., 53, 928; DOI: 10.1016/j.asr.2013.08.024
\bibitem[Bianchi \& Shiao(2020)]{guvmatch20}Bianchi, L. \& Shiao, B. 2020, \apjs, 250:36; DOI: https://doi.org/10.3847/1538-4365/aba2d7 ; arXiv:2007.03808 
\bibitem[Bressan et al.(2012)]{bressan12}Bressan, A., Marigo, P., Girardi, L.  et al. 2012, \mnras, 427, 127
\bibitem[Castelli \& Kurucz(2003)]{castalfa}Castelli, F. \& Kurucz, R.L.\  2003, IAU Symp.210, P.A20C, arXiv: arXiv: arXiv:astro-ph/0405087;  DOI: 10.48550/arXiv.astro-ph/0405087 
\bibitem[D'Orazio \& Samsing(2018)]{dorazio18}D'Orazio, D.J., and Samsing, J.\ 2018, \mnras,481, p.4775 
DOI: 10.1093/mnras/sty2568 
\bibitem[El-Badry et al.(2022)]{elbadry22}El-Badry, K., Seeburger, R., Jayasinghe, T., et al.\ 2022, \mnras, 512, 5620 
\bibitem[El-Badry et al.(2024)]{elbadry24}El-Badry, K., et al.\ 2024, in preparation  
\bibitem[Hubeny \& Long(2021)]{HubenyLong21}Hubeny, I., \& Long, K.S.  2021, \mnras, MNRAS.tmp..821H, arXiv:2103.09341    
\bibitem[Langer et al.(2020)]{Langer2020}Langer, N., Schürmann, C.,  Stoll, K.,  et al. 2020, \aap, 638, 39
\bibitem[Moe \& di~Stefano(2017)]{MoediStefano2017}Moe, M. \& di~Stefano, R.,  2017 \apjs, 230, 15
\bibitem[Ozel et al.(2012)]{Ozel2012}Ozel, F., Psaltis, D., Narayan, R., \& Santos Villarreal, A.\ 2012, \apj,  757, 550
\bibitem[Ozel et al.(2010)]{Ozel2010}Ozel, F., Psaltis, D., Narayan, R., \& McClintock, Jeffrey E.\ 2010 \apj,  725 , 1918
\bibitem[Patrick et al.(2019)]{Patrick2019}Patrick, L.R.,  Lennon, D. J.; Britavskiy, N., et al. 2019 \aap 624, 129
\bibitem[Postma et al.(2021)]{Postmauvit}Postma, J.E. \& Leahy, D., 2021, Journal of Astrophysics and Astronomy, Volume 42, Issue 2, article id.30, DOI: 
10.1007/s12036-020-09689-w 
\bibitem[Postnov \& Yungelson(2006)]{PostnovYungelson2006}Postnov, K.A., \& Yungelson, L.R., 2006, Living Reviews in Relativity, Volume 9, Issue 1, article id. 6, DOI:  10.12942/lrr-2006-6
\bibitem[Puebla et al.(2007)]{puebla07}Puebla, R., Diaz, M.P., \& Hubeny, I.\ 2007, \aj, 134, 1923  
\bibitem[Sana et al.(2014)]{Sanaetal2014}Sana, H., Bouquin, J.B., Lacour, S., et al. 2014, \apjs, 215, 15 
\bibitem[Sana et al.(2017)]{Sanaetal2017}Sana, H., Ramirez-Tannus, M.C., de~Koter, A.,  et al.\ 2017 \aap 599, L9
\bibitem[Strassmeier(2009)]{Strassmeier2009}Strassmeier, K.G. 2009, Astron. Astrophys. Rev., 17, 251-308 
\bibitem[Tandon et al.(2020)]{Tandon2020}Tandon, S. N., Postma, J., Joseph, P., et al. 2020 \aj 159, 158, DOI: 
10.3847/1538-3881/ab72a3  andf 10.48550/arXiv.2002.01159 
\bibitem[Thompson et al.(2019)]{Thompsonetal2019}Thompson,T., Kochanek, C., Stanek, K., et al. 2019, Science 366, 637 ( T2019 )
\bibitem[Thompson et al.(2020)]{Thompsonetal2020}Thompson, T.,  Kochanek, C., Stanek, K., et al. 2020, Science, Volume 368, Issue 6491, id. eaba4356 (2020), DOI: 10.1126/science.aba4356 (T2020)
\bibitem[van den Heuvel \& Tauris(2020)]{vandenHeuvelTauris2020}van den Heuvel, E.P, \& Tauris, T. M. 2020, Science 10.1126/science.aba3282
\end{thebibliography}
\end{document}